%% file: main.tex
\documentclass[letterpaper]{article}

\usepackage[numbers,sort&compress]{natbib}

\usepackage[T1]{fontenc}
\usepackage{graphicx}
\usepackage{microtype}
\usepackage{booktabs}
\usepackage{float}
\usepackage{setspace}
\usepackage{amsmath, amssymb, mathtools}
\allowdisplaybreaks
\usepackage{tikz}
\usetikzlibrary{shapes,decorations,arrows,calc,arrows.meta,fit,positioning}

\usepackage[margin=1in]{geometry}

\usepackage{authblk}

\usepackage{caption}
\usepackage[colorlinks=true,
            linkcolor=red,
            citecolor=red,
            urlcolor=red,
            breaklinks=true]{hyperref}

\newcommand\independent{\protect\mathpalette{\protect\independenT}{\perp}}
\def\independenT#1#2{\mathrel{\rlap{$#1#2$}\mkern2mu{#1#2}}}

\tikzstyle{box} = [rectangle, draw, minimum width=1cm, minimum height=0.7cm, text centered]
\tikzstyle{arrow} = [thick,->,>=stealth]

\usepackage{placeins}
\usepackage{xcolor}
\usepackage{etoolbox}
\usepackage{array,tabularx}

\begin{document}
\title{Causal Variance Decompositions for Measuring Health Inequalities}
\author[1]{Lin Yu}
\author[1,2]{Zhihui (Amy) Liu}
\author[3]{Kathy Han}
\author[1]{Olli Saarela\thanks{Corresponding author Olli Saarela, Dalla Lana School of Public Health, 155 College Street, Toronto, Ontario M5T 3M7, Canada. \href{mailto:olli.saarela@utoronto.ca}{olli.saarela@utoronto.ca}.}}
\affil[1]{Dalla Lana School of Public Health, University of Toronto, Ontario, Canada}
\affil[2]{Department of Biostatistics, Princess Margaret Cancer Centre, University Health Network, Ontario, Canada}
\affil[3]{Radiation Medicine Program, Princess Margaret Cancer Centre, University Health Network, Ontario, Canada}
\date{}
\fontsize{12pt}{12pt}\selectfont
\maketitle
\begin{abstract}
Recent causal inference literature has introduced causal effect decompositions to quantify sources of observed inequalities or disparities in outcomes, but these approaches are typically limited to pairwise comparisons. In healthcare delivery studies, both the care-setting variable (e.g., healthcare provider, institution, or geographic area) and sociodemographic group membership may be polytomous, making pairwise contrasts inadequate. We therefore take the variance of care delivery outcomes under a prespecified stochastic assignment intervention as the quantity of interest and develop a new causal variance decomposition framework for this setting. The proposed framework decomposes this interventional variance into eight components, including novel terms capturing contributions associated with group differences in care-setting distributions, heterogeneity in average care-setting effects across sociodemographic groups, and the covariance between pairwise assignment weights and squared care-setting effects. We discuss the causal interpretation of these components, propose both parametric and nonparametric model-based estimators, and study their performance through simulation. Finally, we illustrate the method using data from the SEER program in an application to cervical cancer care delivery.
\end{abstract}

\section{Introduction}\label{sec:intro}
Marginalized sociodemographic groups may experience disparities in both access to care and the quality of care processes they receive, even in publicly funded systems \cite{williams1995us,williams2010race,price2013racial,canedo2018racial,helpman2020disparities,wolfstadt2019association}. Causal approaches to healthcare quality comparison often estimate average outcome contrasts between care settings, such as healthcare providers, institutions, healthcare systems or regional care environments, after adjustment for patient case-mix \cite{van2024between,varewyck2014shrinkage,daignault2017doubly,daignault2019causal,silber2010hospital,silber2016improving,george2017mortality,asch2021variation}. While useful for summarizing overall differences between settings, these approaches provide limited insight into whether sociodemographic groups are distributed differently across settings (distributional differences) or whether outcome contrasts between settings differ across groups (effect modification). Methods are therefore needed that move beyond average comparisons to distinguish how these care setting distributional differences and effect heterogeneity contribute to differences in outcomes.

Health inequalities or disparities across sociodemographic groups are challenging to study within the traditional exposure–outcome paradigm in causal inference \cite{vanderweele2014causal,jackson2021meaningful,shen2025calibrated}. This difficulty arises when the sociodemographic characteristics, such as race or ethnicity, are not manipulable, so indexing potential outcomes by such an `exposure' is not meaningful. To address this challenge, \textit{causal effect decomposition} methods define potential outcomes on a modifiable/intervenable mediator along the disparity pathway \cite{vanderweele2014causal,jackson2021meaningful}. For example, Jackson and VanderWeele \cite{jackson2018decomposition} proposed defining potential outcomes only with respect to an intervenable mediator and decomposing the observed disparity between groups into two components: (i) the disparity reduction, i.e., the portion eliminated under a hypothetical intervention that equalizes the mediator/treatment distribution across groups; and (ii) the residual disparity, i.e., the portion that remains after such an intervention. The related counterfactual disparity measure proposed by Naimi et al. \cite{naimi2016mediation} captures the portion of the disparity that would remain if the mediator were set to a common reference value across groups. More recent work extends this framework to a four\nobreakdash-way decomposition that attributes disparities to baseline, prevalence, effect modification, and differential selection pathways \cite{yu2025nonparametric}. The baseline component coincides with the counterfactual disparity measure defined by Naimi et al. \cite{naimi2016mediation} and the other three components are measures for different sources of health inequalities. Although these approaches require weaker assumptions than traditional mediation analysis, they are inherently specified for pairwise sociodemographic group or treatment contrasts, limiting their applicability in settings with polytomous treatments and multiple sociodemographic groups, as commonly encountered in provider profiling.

Variance can also serve as an effect measure, and variance\nobreakdash-explained metrics can summarize a potentially large number of pairwise contrasts into a single quantity, thereby avoiding the proliferation of pairwise comparisons inherent in effect\nobreakdash-decomposition approaches. Building on this idea, Chen et al. \cite{chen2020causal,chen2023hierarchical} introduced a causal variance decomposition that attributes observed outcome variation to hospital performance, patient case-mix covariates, and residual variation. However, this decomposition does not separately quantify differences in care-setting distributions across sociodemographic groups or heterogeneity in care-setting effects across these groups and therefore cannot isolate sources of health inequalities. Subsequent work by Chen et al. \cite{chen2022causal} combined causal variance decomposition with mediation analysis to partition outcome variation into components operating through a specified process of care and those operating through all other pathways. While this extension separates mediating pathways, it requires defining potential outcomes for both the exposure and the mediator and does not accommodate settings with exposure\nobreakdash-induced mediator–outcome confounding. Such confounding arises in the present setting when case-mix variables are influenced by the exposure and simultaneously confound the mediator–outcome relationship, making standard natural direct and indirect effects non\nobreakdash-identifiable \cite{vanderweele2014effect}.

Applying these variance-based approaches to study health inequalities also requires explicit decisions about covariate adjustment. The causal variance decomposition approaches reviewed above generally adjust for the full set of measured case-mix covariates without explicitly considering their allowability. Allowability concerns whether adjustment for group differences in a covariate is justified on clinical and equity grounds when defining the disparity of interest \cite{jackson2021meaningful,chang2024importance}. These choices determine which group differences are excluded from the disparity being studied. For example, differences in socioeconomic characteristics such as income may themselves reflect sources of inequality and should not automatically be adjusted away. Although recent causal effect decomposition methods explicitly incorporate allowability designations \cite{jackson2021meaningful, chang2024importance, benmichael2025estimating}, to our knowledge these considerations have not been explicitly incorporated into causal variance decomposition approaches.

Parallel developments in the machine learning literature have measured variable importance via variance explained type quantities. For example, Khan et al. \cite{khan2025marginal} show equivalence between a permutation-based conditional variable importance metric and components in a causal variance decomposition, and express these in terms of the conditional average treatment effect (CATE). Related work has used variable importance metrics to explore how covariates contribute to treatment effect heterogeneity. For instance, Shin et al.  \cite{shin2025treatment} proposed a variable importance measure for multivariate continuous treatments that attributes overall treatment effect heterogeneity to individual covariates. Hines et al. \cite{hines2025variable} develop a loss\nobreakdash-based decomposition of the mean\nobreakdash-squared error of CATE estimation to quantify how specific covariates drive treatment\nobreakdash-effect variation. These approaches share our conceptual interest in identifying the sources of heterogeneity. However, in contrast our goal is to simultaneously identify and quantify sources of outcome variation,  with particular attention to how care setting distributions and effects vary across sociodemographic groups.

In this work, we develop a causal variance decomposition framework for studying health inequalities across sociodemographic groups and care settings. Building on prior variance-based decomposition approaches, the framework incorporates explicit allowability designations and decomposes outcome variance under a specified stochastic assignment intervention. Our framework distinguishes between-group variation associated with differences in care-setting distributions from remaining variation, and separates average care-setting contrasts from heterogeneity in these contrasts across groups. By summarizing multiple group and care-setting contrasts on a common variance scale, the proposed framework provides an overall perspective on disparity-related outcome variation in polytomous settings, with potential applications to healthcare quality assessment, such as comparisons across healthcare providers, healthcare systems, and healthcare regions.

The remainder of this article is organized as follows. Section 2 defines the interventional outcome variance, presents the proposed decomposition and its interpretation, derives identification results under stated causal assumptions, and develops model-based estimators. Section 3 evaluates the estimators in a simulation study. Section 4 illustrates the framework using SEER data on brachytherapy receipt. Section 5 discusses the findings, limitations, and directions for future research.

\section{Causal Variance Decomposition}\label{sec:eight-way}

\subsection{Setup, Notation and Identifying Assumptions}\label{sec:notation}
Let \(A \in \{1,\ldots,J\}\) denote assignment to a care setting or context. Depending on the application, $A$ may be the healthcare provider or institution delivering care, or a broader context such as a healthcare system or geographic area, which can reflect regional differences in healthcare. Let \(Z \in \{1,\ldots,K\}\) denote categorical sociodemographic group membership.
We consider \(J,K\geq 2\), with particular interest in polytomous settings where \(J>2\) or \(K>2\). The dichotomous case \(J=K=2\) is included as a special case.
Let \(Y\) denote a process-of-care outcome, such as the type of treatment received by a patient, and let \(Y(a)\) denote the potential outcome under care-setting assignment \(A=a\).
The potential-outcome formulation provides a framework for studying outcome variation across care settings or contexts. Defining outcomes on \(A\) does not imply that interventions are aimed at moving patients from one care setting or context to another (e.g., between hospitals or geographic areas). The aim is to understand sources of variation in the care process under study across care settings and sociodemographic groups and to identify care processes, treatment delivery and service availability as potential targets for intervention.

For the measured covariates, we incorporate allowability designations that reflect equity considerations \citep{jackson2021meaningful,chang2024importance,jackson2026target}. A covariate is outcome-allowable if adjustment for its group differences is considered appropriate when defining the outcome disparity. It is treatment-allowable if the hypothetical assignment is permitted to depend on that covariate. In this paper, we distinguish two sets of covariates according to their roles in defining the disparity of interest and specifying the hypothetical intervention.
Let \(\boldsymbol X\) denote covariates that are both outcome-allowable and treatment-allowable, which we refer to as fully allowable covariates, or simply allowable covariates. These may include measures of disease progression or clinical indications for treatment for which adjustment is justified for both purposes. Let \(\boldsymbol N\) denote covariates that are neither outcome-allowable nor treatment-allowable, which we refer to as fully non-allowable covariates. Group differences in these are treated as part of the disparity and retained as part of the contrasts. For example, income may be designated as non-allowable when adjustment would remove socioeconomic or structural mechanisms relevant to the measured inequality.
An additional designation, in which covariates are treatment-allowable but not outcome-allowable, may be relevant in some applications \cite{jackson2021meaningful}.
However, for the process-type outcomes considered in this paper, we restrict attention to the two-set formulation and discuss this extension further in Section~\ref{sec:discussion}.

The hypothesized causal relationships among the observed variables
\((A,Z,\boldsymbol X,\boldsymbol N,Y)\) are represented by the directed acyclic graph in Figure~\ref{fig:DAG_include_U}(a), adapted from \citet{naimi2016mediation} to the healthcare comparison setting. Panels (b)--(c) extend this graph to include assignment-related and outcome-related unmeasured characteristics.

\input{Figures/DAG_include_U}
\FloatBarrier

Furthermore, let \(\mathcal G\) denote a hypothetical stochastic intervention on care-setting assignment. Under \(\mathcal G\), an individual
in group \(z\) with allowable covariate profile \(\boldsymbol x\)
receives a random care-setting assignment, denoted by
\(G(z,\boldsymbol x)\), with probabilities $P\{G(z,\boldsymbol x)=a\}
= P(A=a\mid Z=z,\boldsymbol X=\boldsymbol x)$. Within each \((Z,\boldsymbol X)\) stratum, this draw is independent of \(\boldsymbol N\) and the potential outcomes. Thus, this hypothetical intervention preserves the observed assignment distribution within each group and allowable covariate profile, but removes its dependence on \(\boldsymbol N\), consistent with our allowability designations. Let \(Y^{\mathcal G}=Y\{G(Z,\boldsymbol X)\}\) denote the potential outcome under \(\mathcal G\).

We assume (i) counterfactual consistency, \(Y=Y(A)\);
(ii) positivity,
\(P(A=a\mid Z=z,\boldsymbol N=\boldsymbol n,\boldsymbol X=\boldsymbol x)>0\),
for all settings and covariate profiles required by the comparisons; and
(iii) conditional exchangeability,
$Y(a)\perp\!\!\!\perp A\mid(Z,\boldsymbol X,\boldsymbol N)$. Under these assumptions, for all \((a,z,\boldsymbol x)\) on the relevant support, we have
\begin{align*}
E\{Y(a)\mid Z=z,\boldsymbol X=\boldsymbol x\}
&=
\int
E(Y\mid A=a,Z=z,\boldsymbol N=\boldsymbol n,
\boldsymbol X=\boldsymbol x)
\,dF(\boldsymbol n\mid z,\boldsymbol x).
\end{align*}

\subsection{Causal Decomposition of \(V[Y^{\mathcal G}]\)}

We decompose the outcome variation under the prespecified stochastic care-setting intervention \(\mathcal G\), denoted by \(V[Y^{\mathcal G}]\).
We refer to this quantity as the interventional outcome variance.
We start from \(V(Y^{\mathcal G})\) because our primary interest is variation across sociodemographic groups and care settings under a specified stochastic assignment regime. Within each \((Z,\boldsymbol X)\) stratum, this regime preserves the observed marginal distribution of \(A\) while drawing care settings independently of individual values of \(\boldsymbol N\). It therefore compares outcomes across settings using the same natural distribution of \(\boldsymbol N\) within that stratum. Starting with this interventional outcome variance yields a more focused decomposition of covariate, group, setting, and residual variation.
A decomposition beginning with \(V(Y)\) is also possible, but requires additional terms to account for the change from observed to stochastic assignment and its covariance with the variation remaining under that regime. We present the decomposition of observed variance \(V(Y)\) and its relationship with \(V(Y^{\mathcal G})\) in Appendix~\ref{app:observed-variance-N}.

Applying the law of total variance successively to \(V(Y^{\mathcal G})\) according to the attribution ordering
\(\boldsymbol X\rightarrow Z\rightarrow G\) gives
\begin{align}
V(Y^{\mathcal G})
&=
V_{\boldsymbol X}
+V_{\mathrm{group}}
+V_{\mathrm{setting}}
+V_{\mathrm{residual}},
\label{eq:overall_interventional_decomposition}
\end{align}
where
\begin{align*}
V_{\boldsymbol X}
&=
V_{\boldsymbol X}
\left\{E(Y^{\mathcal G}\mid\boldsymbol X)\right\},\\
V_{\mathrm{group}}
&=
E_{\boldsymbol X}
\left[
V_{Z\mid\boldsymbol X}
\left\{E(Y^{\mathcal G}\mid\boldsymbol X,Z)\right\}
\right],\\
V_{\mathrm{setting}}
&=
E_{\boldsymbol X,Z}
\left[
V_{G\mid\boldsymbol X,Z}
\left\{E(Y^{\mathcal G}\mid\boldsymbol X,Z,G)\right\}
\right],\\
V_{\mathrm{residual}}
&=
E_{\boldsymbol X,Z,G}
\left[
V(Y^{\mathcal G}\mid\boldsymbol X,Z,G)
\right].
\end{align*}
Together, \(V_{\mathrm{group}}\), \(V_{\mathrm{setting}}\), and \(V_{\mathrm{residual}}\) sum to the average interventional outcome variance within allowable covariate profiles, \(E_{\boldsymbol X}\{V(Y^{\mathcal G}\mid\boldsymbol X)\}\), which could also have been taken directly as the quantity being decomposed.

Since we decompose the interventional outcome variance, all components describe outcome variation under the prespecified stochastic intervention.
This intervention is implicit in the interpretations below unless otherwise stated. Here, \(V_{\boldsymbol X}\) represents variation associated with \(\boldsymbol X\).
\(V_{\mathrm{group}}\) represents average between-group variation in expected outcomes conditional on \(\boldsymbol X\).
When \(\boldsymbol N\) is absent, this component reduces, under the identification assumptions, to the observed between-group variation conditional on \(\boldsymbol X\).
\(V_{\mathrm{setting}}\) represents average variation across care settings conditional on \((Z,\boldsymbol X)\), with setting contrasts interpreted causally under the identification assumptions.
\(V_{\mathrm{residual}}\) captures the remaining outcome variation within \((G,Z,\boldsymbol X)\) strata.

\subsubsection{Decomposition of \(V_{\mathrm{group}}\)}
\label{sec:v_group}

For discrete \(Z\), \(V_{\mathrm{group}}\) can be written as
\begin{align}
V_{\mathrm{group}}
&=
E_{\boldsymbol X}
\left\{
\sum_z
\delta_{\mathrm{group}}(z,\boldsymbol X)^2
P(z\mid\boldsymbol X)
\right\},
\label{eq:vgroup_policy}
\end{align}
where
\begin{align}
\delta_{\mathrm{group}}(z,\boldsymbol x)
&=
\sum_a
E\{Y(a)\mid z,\boldsymbol x\}
P(a\mid z,\boldsymbol x)
\notag\\
&\quad-
\sum_{z^*,a}
E\{Y(a)\mid z^*,\boldsymbol x\}
P(a\mid z^*,\boldsymbol x)
P(z^*\mid\boldsymbol x).
\label{eq:group_contrast}
\end{align}

Thus, at each covariate profile \(\boldsymbol x\), each group's expected interventional outcome is compared with the population mean at that profile. Because \(\delta_{\mathrm{group}}(z,\boldsymbol x)\) for all \(z\) is squared before averaging over \(Z\), positive and negative deviations across groups do not cancel in \(V_{\mathrm{group}}\).

We show in Appendix~\ref{sec:V_group_proof} that \(V_{\mathrm{group}}\) can be further decomposed into three terms to distinguish different mechanisms leading to group-related outcome variation. Specifically, we have
\begin{align}
V_{\mathrm{group}}
&=
V_{\mathrm{dist}}
+V_{\mathrm{remain}}
+V_{\mathrm{cross}}
\notag\\
&=
E_{\boldsymbol X}
\left\{
\sum_z
\delta_{\mathrm{dist}}(z,\boldsymbol X)^2
P(z\mid\boldsymbol X)
\right\}
+
E_{\boldsymbol X}
\left\{
\sum_z
\delta_{\mathrm{remain}}(z,\boldsymbol X)^2
P(z\mid\boldsymbol X)
\right\}
\notag\\
&\quad+
2E_{\boldsymbol X}
\left\{
\sum_z
\delta_{\mathrm{dist}}(z,\boldsymbol X)
\delta_{\mathrm{remain}}(z,\boldsymbol X)
P(z\mid\boldsymbol X)
\right\}.
\label{eq:vgroup_three_parts}
\end{align}

This decomposition is expressed in terms of the
\emph{care-setting distribution contrast}
\(\delta_{\mathrm{dist}}(z,\boldsymbol x)\)
and the
\emph{remaining group contrast}
\(\delta_{\mathrm{remain}}(z,\boldsymbol x)\),
defined as
\begin{align}
\delta_{\mathrm{dist}}(z,\boldsymbol x)
&=
\sum_a
E\{Y(a)\mid z,\boldsymbol x\}
\left\{
P(a\mid z,\boldsymbol x)
-P(a\mid\boldsymbol x)
\right\},
\label{eq:delta_dist_N}
\end{align}
and
\begin{align}
\delta_{\mathrm{remain}}(z,\boldsymbol x)
&=
\sum_a
E\{Y(a)\mid z,\boldsymbol x\}
P(a\mid\boldsymbol x)
\notag\\
&\quad-
\sum_{z^*,a}
E\{Y(a)\mid z^*,\boldsymbol x\}
P(a\mid z^*,\boldsymbol x)
P(z^*\mid\boldsymbol x).
\label{eq:delta_remain_N}
\end{align}
Here, \(P(a\mid\boldsymbol x)\) defines the prevalence-weighted
care-setting assignment distribution, obtained by averaging
the group-specific assignment probabilities:
\[
P(a\mid\boldsymbol x)
=
\sum_{z^*}
P(a\mid z^*,\boldsymbol x)
P(z^*\mid\boldsymbol x).
\]
This distribution serves as a common reference \(A\) distribution.

For a given \(\boldsymbol x\),
\(\delta_{\mathrm{dist}}(z,\boldsymbol x)\)
compares the expected outcomes of the same group \(z\) under two stochastic interventions: one randomly assigning care settings according to \(P(a\mid z,\boldsymbol x)\), and the other according to
\(P(a\mid\boldsymbol x)\).
Under both interventions, each group's natural distribution of \(\boldsymbol N\) conditional on \(\boldsymbol X\) is retained.
\(V_{\mathrm{dist}}\) averages these squared contrasts across groups and allowable covariates, summarizing the magnitude of the expected outcome differences associated with replacing group-specific assignment distributions with the common reference distribution.
Squaring each contrast before averaging prevents positive and negative contrasts from cancelling across groups.

For a given \(\boldsymbol x\),
\(\delta_{\mathrm{remain}}(z,\boldsymbol x)\)
compares group \(z\)'s expected outcome with \(A\) randomly drawn from the common reference distribution,
with the population mean under the original group-specific stochastic interventions.
\(V_{\mathrm{remain}}\) averages these squared contrasts across groups and covariate profiles, summarizing deviations from this population benchmark after assignment standardization.

To illustrate these contrasts, consider the case where both \(A\) and \(Z\) are dichotomous, with
\(A\in\{0,1\}\) and \(Z\in\{z,z^*\}\),
where \(z\) denotes the index group and \(z^*\) the comparison group.  Appendix ~\ref{sec:binary_group_contrasts}
derives the expressions below for both contrasts. For the index group \(z\),
\(\delta_{\mathrm{dist}}(z,\boldsymbol x)\)
takes the form
\begin{align*}
\delta_{\mathrm{dist}}(z,\boldsymbol x)
&=
P(Z=z^*\mid\boldsymbol x)
E\{Y(1)-Y(0)\mid z,\boldsymbol x\}\\
&\quad\times
\left[
P(A=1\mid z,\boldsymbol x)
-
P(A=1\mid z^*,\boldsymbol x)
\right].
\end{align*}
In the dichotomous case,  $\delta_{\mathrm{dist}}(z,\boldsymbol x)$ has a structure similar to the prevalence component of Yu and Elwert \cite{yu2025nonparametric}. It compares the expected outcome for members of group \(z\) (given $\boldsymbol x$) under their own assignment distribution with their expected outcome under the assignment distribution of group \(z^*\) (given $\boldsymbol x$). The contrast is weighted by \(P(Z=z^*\mid\boldsymbol x)\) because this is the share of the common reference distribution
that comes from group \(z^*\). For \(\delta_{\mathrm{dist}}(z,\boldsymbol x)\) to be nonzero, both the average outcome contrast between \(A=1\) and \(A=0\) and the assignment-probability contrast must be nonzero.

The remaining group contrast simplifies to
\[
\begin{aligned}
\delta_{\mathrm{remain}}(z,\boldsymbol x)
&=
P(Z=z^*\mid\boldsymbol x)\Bigl\{
E\{Y(0)\mid z,\boldsymbol x\}
-E\{Y(0)\mid z^*,\boldsymbol x\}\\
&\quad {}+
P(A=1\mid Z=z^*,\boldsymbol x)\\
&\qquad {}\times \Bigl[
E\{Y(1)-Y(0)\mid z,\boldsymbol x\}
-E\{Y(1)-Y(0)\mid z^*,\boldsymbol x\}
\Bigr]
\Bigr\}.
\end{aligned}
\]
The two terms inside the braces are algebraically analogous to the baseline and effect components of the disparity decomposition of \citet{yu2025nonparametric},
although the reference and weighting choices differ.
This connection highlights that the remaining group contrast reflects both baseline differences in expected outcomes under \(A=0\) and differences in conditional average setting effects across groups, with the latter weighted by \(P(A=1\mid Z=z^*,\boldsymbol x)\).

In the general polytomous case, Appendix~\ref{sec:V_remain_further} gives two reference choices for splitting the remaining group contrast into baseline and effect (modification) contrasts. We retain \(\delta_{\mathrm{remain}}\) as a single contrast because this split would introduce three pairwise cross terms among the distribution, baseline, and effect contrasts in \(V_{\mathrm{group}}\). The resulting effect modification component generally has no variance-of-effects interpretation, whereas \(V_{\mathrm{em}}\) in Section~\ref{sec:v_hospital} directly summarizes the variance across groups of pairwise care-setting effects at a given allowable covariate profile.

\(V_{\mathrm{cross}}\) captures whether the care-setting distribution and remaining group contrasts reinforce or offset one another. In the dichotomous case \(Z\in\{z,z^*\}\), we have
\[
\begin{aligned}
V_{\mathrm{cross}}
=
2E_{\boldsymbol X}\Bigl\{&
P(Z=z\mid\boldsymbol X)
\delta_{\mathrm{dist}}(z,\boldsymbol X)
\delta_{\mathrm{remain}}(z,\boldsymbol X)\\
&+
P(Z=z^*\mid\boldsymbol X)
\delta_{\mathrm{dist}}(z^*,\boldsymbol X)
\delta_{\mathrm{remain}}(z^*,\boldsymbol X)
\Bigr\}.
\end{aligned}
\]
At each allowable covariate profile, the products of the two contrasts (i.e., $\delta_{\mathrm{dist}}(z,\boldsymbol X)$ and $\delta_{\mathrm{remain}}(z,\boldsymbol X)$) are averaged across groups using the conditional group probabilities as weights. \(V_{\mathrm{cross}}\) is twice the average of these weighted means over \(\boldsymbol X\), and is not generally the covariance. Within a given group and covariate profile, the two contrasts reinforce one another when they have the same sign and offset one another when they have opposite signs. Positive and negative products can cancel across groups and covariate profiles. Thus,
a positive \(V_{\mathrm{cross}}\) indicates that reinforcement dominates on average, and a negative value indicates that offsetting dominates.  \(V_{\mathrm{cross}}\) equals twice the covariance between the two contrasts across groups, averaged over \(\boldsymbol X\) in the special case of the absence of $A$ and $Z$ interaction (see Appendix~\ref{sec:V_cross}).

Because potential outcomes in our formulation are indexed only by
care-setting assignment \(A\), the group-related components are formulated as associational and do not
isolate distinct causal pathways from \(Z\).
Appendix~\ref{sec:path_specific_effect} establishes a path-specific
causal interpretation of the group contrasts in a special case in which
potential outcomes are also defined with respect to \(Z\) and
stronger identification assumptions hold. In the general
formulation, these components summarize group comparisons under
the specified assignment interventions. These comparisons may
also reflect selection into covariate strata in the presence of
unmeasured characteristics, as illustrated in panels (b) and (c)
of Figure~\ref{fig:DAG_include_U}. In these examples, conditioning
on \(\boldsymbol X\) can induce associations between \(Z\) and
\(U_1\) or \(U_2\), which may contribute to group differences in
assignment distributions and conditional potential-outcome means.
Following \citet{jackson2026target}, whether to treat these
selection-related differences as bias or retain them in the
comparison depends on the target population and the disparity of
interest. Our framework retains these differences in comparisons
across observed groups within \(\boldsymbol X\) strata in the
study population.

To illustrate how these selection-related differences enter the
group contrasts, we consider \(U_1\) and \(U_2\) separately, using setting-specific potential-outcome means averaged within each \((Z,\boldsymbol X)\) stratum. If \(U_1\) and \(U_2\) were both present in the same DAG, conditioning on \((Z,\boldsymbol X,\boldsymbol N)\) would open backdoor paths such as \(A\leftarrow U_1\to\boldsymbol X\leftarrow U_2\to Y\), so conditional exchangeability could fail. We therefore focus on panels (b) and (c), where conditional exchangeability holds given the same adjustment set.

For assignment-related selection through \(U_1\) in Figure~\ref{fig:DAG_include_U} panel (b), we examine the assignment probabilities while holding these means fixed. The contribution through the assignment probabilities is then confined to \(\delta_{\mathrm{dist}}\) in the absence of an additive \(A\times Z\) interaction in the averaged means and may enter both contrasts when this interaction is present. For outcome-related selection through \(U_2\) in Figure~\ref{fig:DAG_include_U} panel (c), its contribution through the potential-outcome mean under the illustrative additive model is confined to \(\delta_{\mathrm{remain}}\) in the absence of an \(A\times U_2\) interaction and may enter both contrasts when this interaction is present. This result holds regardless of the presence of an \(A\times Z\) interaction. Detailed derivations are provided in Appendix~\ref{sec:unmeasured_u}.

\subsubsection{Decomposition of \(V_{\mathrm{setting}}\)}
\label{sec:v_hospital}

Similar to \(V_{\mathrm{group}}\), the care-setting variation
\(V_{\mathrm{setting}}\) reflects several sources of variation.
As shown in Appendix~\ref{sec:V_setting_proof},
\(V_{\mathrm{setting}}\) can be decomposed into the main effect
component for care setting (\(V_{\mathrm{me}}\)), the component
for effect modification by group (\(V_{\mathrm{em}}\)),
and the covariance component for pairwise assignment weights
and squared setting effects (\(V_{\mathrm{cov}}\)):
\begin{align*}
V_{\mathrm{setting}}
&=
V_{\mathrm{me}}
+V_{\mathrm{em}}
+V_{\mathrm{cov}}\\
&=
E_{\boldsymbol X}
\left\{
\sum_{a<a^*}
E_{Z\mid\boldsymbol X}
\{\pi_{aa^*}(Z,\boldsymbol X)\}
\left[
E_{Z\mid\boldsymbol X}
\{\delta_{aa^*}(Z,\boldsymbol X)\}
\right]^2
\right\}\\
&\quad+
E_{\boldsymbol X}
\left\{
\sum_{a<a^*}
E_{Z\mid\boldsymbol X}
\{\pi_{aa^*}(Z,\boldsymbol X)\}
V_{Z\mid\boldsymbol X}
\{\delta_{aa^*}(Z,\boldsymbol X)\}
\right\}\\
&\quad+
E_{\boldsymbol X}
\left\{
\sum_{a<a^*}
\operatorname{Cov}_{Z\mid\boldsymbol X}
\left(
\pi_{aa^*}(Z,\boldsymbol X),
\delta_{aa^*}(Z,\boldsymbol X)^2
\right)
\right\},
\end{align*}
where, for each unordered care-setting pair \((a,a^*)\), define
\begin{align*}
\pi_{aa^*}(z,\boldsymbol x)
&=
P(A=a\mid z,\boldsymbol x)
P(A=a^*\mid z,\boldsymbol x),\\
\delta_{aa^*}(z,\boldsymbol x)
&=
E\{Y(a)\mid z,\boldsymbol x\}
-E\{Y(a^*)\mid z,\boldsymbol x\}.
\end{align*}
The product \(\pi_{aa^*}(z,\boldsymbol x)\) is the pairwise weight  for setting pair (\(a\), \(a^*\)) in the pairwise representation of a categorical variance. It is not the probability that the same individual is assigned to both settings and does not require independence between two visits.
The contrast \(\delta_{aa^*}(z,\boldsymbol x)\) is a conditional average causal effect of assignment to \(a\) rather than \(a^*\). Each component is defined at the level of a care-setting pair and then averaged over \(\boldsymbol X\). For polytomous \(A\), the contributions are summed over all unordered pairs.

The main effect component \(V_{\mathrm{me}}\) captures variation associated with the average difference in outcomes between care settings and may be relevant to group inequalities through differences in care-setting assignment.
In the dichotomous case \(A\in\{0,1\}\),
\(V_{\mathrm{me}}\) simplifies to
\begin{align*}
V_{\mathrm{me}}
&=
E_{\boldsymbol X}\Bigl\{
E_{Z\mid\boldsymbol X}\left[
P(A=1\mid Z,\boldsymbol X)
\left\{1-P(A=1\mid Z,\boldsymbol X)\right\}
\right]\\
&\qquad {}\times
\left[
E\{Y(1)\mid\boldsymbol X\}
-
E\{Y(0)\mid\boldsymbol X\}
\right]^2
\Bigr\}.
\end{align*}
In the dichotomous $A$ case, the pairwise weight
\[
\pi_{01}(z,\boldsymbol x) = P(A=1\mid z,\boldsymbol x)
\left\{1-P(A=1\mid z,\boldsymbol x)\right\},
\]
which is a conditional Bernoulli variance and has the same form as the tilting function used in propensity score overlap weighting \citep{li2018balancing}.  $E_{Z\mid\boldsymbol X}\left[
P(A=1\mid Z,\boldsymbol X)
\left\{1-P(A=1\mid Z,\boldsymbol X)\right\}
\right]$ then summarizes,  on average across sociodemographic groups, the representation (or overlap) of both care settings among patients with allowable covariate profile \(\boldsymbol X\). The squared term represents the squared average causal contrast between settings \(A=1\) and \(A=0\) for patients with allowable covariate profile \(\boldsymbol X\). This squared setting effect is weighted by this averaged overlap weight. Intuitively, even if the average effect comparing \(A=1\) and \(A=0\) is large, its contribution to interventional outcome variation is scaled down when either setting has a low assignment probability within the relevant group and covariate strata.
Because each pairwise average setting contrast is squared, when $A$ is polytomous, summing across pairs does not cancel positive and negative average contrasts, allowing this component to summarize average care-setting effects across all setting pairs.

\(V_{\mathrm{em}}\) captures variation associated with differences in main care-setting effects across sociodemographic groups.
In the dichotomous case \(A\in\{0,1\}\), we have
\begin{align*}
V_{\mathrm{em}}
&=
E_{\boldsymbol X}\Bigl\{
E_{Z\mid\boldsymbol X}\left[
P(A=1\mid Z,\boldsymbol X)
\left\{1-P(A=1\mid Z,\boldsymbol X)\right\}
\right]\\
&\qquad {}\times
V_{Z\mid\boldsymbol X}\left[
E\{Y(1)\mid Z,\boldsymbol X\}
-
E\{Y(0)\mid Z,\boldsymbol X\}
\right]
\Bigr\}.
\end{align*}
Here, the \(V_{Z\mid\boldsymbol X}[\cdot]\) term measures whether the average causal effect comparing settings \(A=1\) and \(A=0\) varies across groups at the same allowable covariate profile. A nonzero value indicates effect heterogeneity across the groups.
Such heterogeneity may arise when setting effects differ by \(Z\) for given
\(\boldsymbol x\), or when setting effects vary with
\(\boldsymbol N\) and its distribution given \(\boldsymbol x\) differs across \(Z\) groups.
The averaged overlap weight
\(E_{Z\mid\boldsymbol X}\{.\}\)
again reflects the representation of the setting pair in the patient population.

The \(V_{\mathrm{cov}}\) component captures a higher-order disparity-relevant contribution to interventional outcome variation arising from the covariance between care-setting assignment overlap and squared setting contrasts.
This term reflects whether the sociodemographic groups for whom the absolute difference between settings is larger also have greater overlap in care setting assignment.
In the dichotomous case \(A\in\{0,1\}\), we obtain

\begin{align*}
V_{\mathrm{cov}}
&=
E_{\boldsymbol X}\Bigl\{
\operatorname{Cov}_{Z\mid\boldsymbol X}\Bigl(
P(A=1\mid Z,\boldsymbol X)
\left\{1-P(A=1\mid Z,\boldsymbol X)\right\},\\
&\qquad \left[
E\{Y(1)\mid Z,\boldsymbol X\}
-
E\{Y(0)\mid Z,\boldsymbol X\}
\right]^2
\Bigr)
\Bigr\}.
\end{align*}

Here, for a given setting pair \((0,1)\) and allowable covariate profile \(\boldsymbol X\), a positive covariance indicates that sociodemographic groups with larger squared setting contrasts also have greater overlap in this setting pair, increasing the pair's contribution to setting-related outcome variation relative to the sum of its main effect and effect modification components.
Conversely, a negative covariance indicates that larger squared contrasts occur in groups with smaller overlap, reducing that contribution. If the setting contrast does not vary across groups, or if the care-setting assignment distribution does not differ across groups, the covariance is zero.

For \(V_{\mathrm{cov}}\), when $A$ is polytomous, the pairwise care-setting contributions can be positive or negative and may cancel one another, so a small overall value does not necessarily imply the absence of this type of covariance within individual setting pairs.
Nonetheless, the aggregate measure summarizes the population-level contribution of covariance between pairwise assignment weights and squared setting contrasts across all setting pairs. And this summary is invariant to relabeling of
the sociodemographic groups.

\subsection{Identification and Estimation}
\label{sec:identification-estimation}

Under the causal assumptions listed in Section~\ref{sec:notation}, the following contrasts can be identified as
\begin{align}
\delta_{\mathrm{dist}}(z,\boldsymbol x)
&=
\sum_{z^*,a}
\left[
\int
E(Y\mid a,z,\boldsymbol n,\boldsymbol x)
\,dF(\boldsymbol n\mid z,\boldsymbol x)
\right]
\notag\\
&\quad\times
\left\{
P(a\mid z,\boldsymbol x)
-P(a\mid z^*,\boldsymbol x)
\right\}
P(z^*\mid\boldsymbol x),
\label{eq:identified-dist}\\
\delta_{\mathrm{remain}}(z,\boldsymbol x)
&=
\sum_{z^*,a}
\Biggl\{
\int
E(Y\mid a,z,\boldsymbol n,\boldsymbol x)
\,dF(\boldsymbol n\mid z,\boldsymbol x)
\notag\\
&\qquad-
\int
E(Y\mid a,z^*,\boldsymbol n,\boldsymbol x)
\,dF(\boldsymbol n\mid z^*,\boldsymbol x)
\Biggr\}
P(a\mid z^*,\boldsymbol x)
P(z^*\mid\boldsymbol x),
\label{eq:identified-remain}\\
\delta_{aa^*}(z,\boldsymbol x)
&=
\int
\left\{
E(Y\mid a,z,\boldsymbol n,\boldsymbol x)
-E(Y\mid a^*,z,\boldsymbol n,\boldsymbol x)
\right\}
\,dF(\boldsymbol n\mid z,\boldsymbol x).
\label{eq:identified-setting-contrast}
\end{align}

We propose model-based plug-in estimators for the variance components. Estimation is based on four nuisance functions and the empirical distribution of $\boldsymbol{X}$.
In \eqref{eq:identified-dist}--\eqref{eq:identified-setting-contrast},
\(E(Y\mid A,Z,\boldsymbol N,\boldsymbol X)\),
\(f(\boldsymbol N\mid Z,\boldsymbol X)\),
and \(P(Z\mid\boldsymbol X)\) are modeled directly.
\(P(A\mid Z,\boldsymbol X)\) is obtained by modeling \(P(A\mid Z,\boldsymbol N,\boldsymbol X)\)
and integrating over \(f(\boldsymbol N\mid Z,\boldsymbol X)\).
While we could model \(P(A\mid Z,\boldsymbol X)\) directly, we choose the former approach as it corresponds to a specific factorization of the likelihood into the four components, which allows us to use an approximate Bayesian procedure described below to evaluate uncertainty in the variance component estimates.

\paragraph*{Outcome model}
\label{sec:outcome_model}

We consider two parametric specifications:
(a) a generalized linear fixed-effects model,
\begin{align*}
g\!\left[
E(Y\mid A,Z,\boldsymbol N,
\boldsymbol X;\boldsymbol\theta)
\right]
&=
\theta_0
+\boldsymbol\theta_1^{T}\boldsymbol X
+\boldsymbol\theta_2^{T}\boldsymbol N\\
&\quad+
\sum_{z=2}^{K}
\theta_3^z\mathbb I(Z=z)
+
\sum_{a=2}^{J}
\theta_4^a\mathbb I(A=a)\\
&\quad+
\sum_{a=2}^{J}
\sum_{z=2}^{K}
\theta_5^{a,z}
\mathbb I(A=a)\mathbb I(Z=z),
\end{align*}
where
\(\boldsymbol\theta=(\theta_0,\allowbreak\boldsymbol\theta_1^{T},\allowbreak
\boldsymbol\theta_2^{T},\allowbreak\theta_3^2,\ldots,\theta_3^K,\allowbreak
\theta_4^2,\ldots,\theta_4^J,\allowbreak
\theta_5^{2,2},\ldots,\theta_5^{J,K})^{T}\),
and \(A=1\) and \(Z=1\) serve as reference levels without loss of generality;
and (b) a generalized linear mixed-effects model with care-setting-specific random intercepts and group slopes,
\begin{align}
g\!\left[
E(Y\mid A,Z,\boldsymbol N,
\boldsymbol X;\boldsymbol\theta,\boldsymbol b)
\right]
&=
\theta_0
+\boldsymbol\theta_1^{T}\boldsymbol X
+\boldsymbol\theta_2^{T}\boldsymbol N
\notag\\
&\quad+
\sum_{z=2}^{K}
\theta_3^z\mathbb I(Z=z)
+
\sum_{a=1}^{J}
b_{0a}\mathbb I(A=a)
\notag\\
&\quad+
\sum_{a=1}^{J}
\sum_{z=2}^{K}
b_{1az}\mathbb I(A=a)\mathbb I(Z=z),
\label{eq:random_effect_est}
\end{align}
where \(b_{0a}\) is the random intercept for
care setting \(a\), and \(b_{1az}\) is the random slope for the contrast
between group \(z\) and the reference group within setting \(a\), for \(z\in\{2,\ldots,K\}\).
Let \(\boldsymbol b_a=(b_{0a},b_{1a2},\ldots,b_{1aK})^{T}\) collect the random effects for setting $a$, and let \(\boldsymbol b=(\boldsymbol b_1^{T},\ldots,\boldsymbol b_J^{T})^{T}\).
The random effects are assumed to have a multivariate normal distribution with zero mean and a
\((J(1+(K-1))\times J(1+(K-1)))\)
covariance matrix taken to be block-diagonal by the levels of \(A\), with common variance and covariance parameters between the groups.

\paragraph*{Assignment models}

We use multinomial logistic models for
\(P(A\mid Z,\boldsymbol N,\boldsymbol X)\)
and \(P(Z\mid\boldsymbol X)\):
\begin{equation}
P(A=a\mid Z,\boldsymbol N,\boldsymbol X;\boldsymbol\eta)
=
\begin{cases}
\displaystyle
\frac{1}
{1+\sum_{j=2}^{J}
\exp\{c_j(Z,\boldsymbol N,\boldsymbol X)\}},
&a=1,\\[10pt]
\displaystyle
\frac{\exp\{c_a(Z,\boldsymbol N,\boldsymbol X)\}}
{1+\sum_{j=2}^{J}
\exp\{c_j(Z,\boldsymbol N,\boldsymbol X)\}},
&a=2,\ldots,J,
\end{cases}
\label{eq:assignment-full-N}
\end{equation}
with
\[
c_a(Z,\boldsymbol N,\boldsymbol X)
=
\eta_{0a}
+\boldsymbol\eta_{Xa}^{\mathsf T}\boldsymbol X
+\sum_{z=2}^{K}\eta_{az}
\mathbb I(Z=z)
+\boldsymbol\eta_{Na}^{\mathsf T}\boldsymbol N,
\]
and
\begin{equation}
P(Z=z\mid\boldsymbol X;\boldsymbol\phi)
=
\begin{cases}
\displaystyle
\frac{1}
{1+\sum_{k=2}^{K}
\exp\{\phi_{0k}+\boldsymbol\phi_k^{\mathsf T}\boldsymbol X\}},
&z=1,\\[10pt]
\displaystyle
\frac{\exp\{\phi_{0z}+\boldsymbol\phi_z^{\mathsf T}\boldsymbol X\}}
{1+\sum_{k=2}^{K}
\exp\{\phi_{0k}+\boldsymbol\phi_k^{\mathsf T}\boldsymbol X\}},
&z=2,\ldots,K.
\end{cases}
\label{eq:group-model}
\end{equation}
We let \(\boldsymbol\eta\) and \(\boldsymbol\phi\) be the vectors of all parameters in the parametrized models
\(P(A\mid Z,\boldsymbol N,\boldsymbol X;\boldsymbol\eta)\)
and
\(P(Z\mid\boldsymbol X;\boldsymbol\phi)\),
respectively.

\paragraph*{Non-allowable covariate model}

For a vector \(\boldsymbol N=(N_1,\ldots,N_L)\), we use sequential conditional models,
\begin{equation}
f(\boldsymbol n\mid z,\boldsymbol x;\boldsymbol\beta)
=
\prod_{\ell=1}^{L}
f(n_\ell\mid n_1,\ldots,n_{\ell-1},z,
\boldsymbol x;\boldsymbol\beta_\ell).
\label{eq:sequential-N}
\end{equation}
The distribution and link for each factor depend on the variable type, for example Gaussian regression for a continuous variable, logistic regression for a binary variable, and ordinal or multinomial logistic regression for a categorical variable. Point estimates of the variance components can be obtained by substituting predictions from the fitted models and standardizing over the observed allowable covariate distribution. The plug-in estimator formulas are provided in Appendix \ref{sec:model_based_estimator}.

\paragraph*{Uncertainty propagation}

We use the Bayesian bootstrap to quantify uncertainty in the variance component estimates. For a dataset with \(n\) patients, each Bayesian bootstrap draw \(b=1,\ldots,B\) assigns a new vector of patient-level weights $(w_1^{(b)},\ldots,w_n^{(b)}) \sim \operatorname{Dirichlet}(1,\ldots,1)$ to those same \(n\) patients. We use these weights to refit all nuisance models and recompute all variance components, averaging over the corresponding weighted empirical distribution of \(\boldsymbol X\). The resulting draws provide an approximate posterior distribution for the variance components. The advantage of the Bayesian over frequentist bootstrap is that the continuous sampling weights don't create further empty categories/separation when re-fitting the models.

For the fixed-effects outcome model, we can also employ parametric approximate Bayesian inference based on the factorized likelihood and normal approximations of the posteriors under block-independent priors for the four parameter vectors. Let
\(\boldsymbol\theta^{(b)}\),
\(\boldsymbol\eta^{(b)}\),
\(\boldsymbol\phi^{(b)}\),
and \(\boldsymbol\beta^{(b)}\)
denote random draws for \(b=1,\ldots,B\) from the multivariate normal distributions $
N\!\left(
\hat{\boldsymbol\theta},
\widehat V(\hat{\boldsymbol\theta})
\right), N\!\left(
\hat{\boldsymbol\eta},
\widehat V(\hat{\boldsymbol\eta})
\right), N\!\left(
\hat{\boldsymbol\phi},
\widehat V(\hat{\boldsymbol\phi})
\right)$ and $N\!\left(
\hat{\boldsymbol\beta},
\widehat V(\hat{\boldsymbol\beta})
\right)$, using the maximum likelihood estimators and their estimated variance--covariance matrices from the corresponding models. For each sample of parameter vectors, we recompute the variance components, and the distribution of these is taken as an approximate posterior distribution of the variance components.

\section{Simulation Study}\label{sec:simulation}

\subsection{Simulation Setup}

We investigated the performance of the proposed estimators
through a simulation study to examine their behavior under
varying sample sizes and numbers of care settings.
The chosen data-generating mechanism was consistent with the DAG in Figure~\ref{fig:DAG_include_U}(a).
We fixed the number of sociodemographic
groups \(Z\in\{1,\ldots,K\}\) at \(K=3\) and generated \(Z\) from
\[
\bigl(\mathbb I(Z=1),\mathbb I(Z=2),\mathbb I(Z=3)\bigr)
\sim\operatorname{Multinomial}(1;\,0.45,0.35,0.20).
\]
We simulated one continuous non-allowable covariate
\(\boldsymbol N=N\) and two allowable covariates, one binary
\(X_1\) and one continuous \(X_2\).
They were generated from the following distributions:
\begin{align}
N\mid Z=z
&\sim
\mathcal N\!\left(
-0.25+0.35\mathbb I(z=2)+0.70\mathbb I(z=3),
\,0.70^2
\right),
\notag\\
\operatorname{logit}P(X_1=1\mid z,n)
&=
-0.10+0.10\mathbb I(z=2)+0.20\mathbb I(z=3)+0.30n,
\notag\\
X_2\mid Z=z,N=n
&\sim
\mathcal N\!\left(
0.60+0.10\mathbb I(z=2)+0.20\mathbb I(z=3)+0.35n,
\,0.55^2
\right).
\label{eq:simulation-covariates}
\end{align}
Care-setting assignment \(A\in\{1,\ldots,J\}\) was generated from a multinomial logistic regression model (Section~\ref{sec:identification-estimation}), conditional on \(Z\), \(N\), \(X_1\), and \(X_2\). We considered two scenarios for the number of hospitals, \(J=5\) and \(J=10\). The outcome was generated from a fixed-effects generalized linear model, conditional on allowable covariates \(X_1\) and \(X_2\), the non-allowable covariate \(N\), care-setting assignment \(A\), and sociodemographic group \(Z\), including \(A\times Z\) and \(A\times N\) interactions. We used \(A=1\) and \(Z=1\) as reference categories in the generating models. The two scenarios shared the coefficients for settings 1 through 5, and the \(J=10\) scenario added coefficients for settings 6 through 10. In addition to varying the number of settings, we also considered both continuous and dichotomous outcomes by applying different link functions, while keeping the same linear-predictor coefficients within each setting scenario. The full data-generating mechanism, including all distributions and coefficients, is provided in Appendix~\ref{sec:simulation_dgm}. These scenarios were further evaluated at sample sizes of 500, 1000, 2500, and 5000 observations.

To compute the true variance component values, we generated a super-population of 10,000 observations for each simulation scenario, using the known parameter values and applied the decomposition to this super-population to obtain Monte Carlo approximations that serve as benchmarks for evaluating estimator performance. For estimator evaluation, we generated 1,000 samples under each
scenario from the same data-generating mechanism. Within each replicate, we fitted a multinomial logistic model for \(A\mid Z,N,\boldsymbol X\), a multinomial logistic model for \(Z\mid\boldsymbol X\), and a Gaussian regression model for \(N\mid Z,\boldsymbol X\). The conditional assignment model matched the generating specification, whereas the group-membership and non-allowable covariate models were working models that need not coincide with the exact conditional distributions implied by the generating mechanism.

The outcome model was estimated using both a fixed-effects model
and a generalized linear mixed-effects model specified in Section \ref{sec:identification-estimation}, the latter being
used to achieve shrinkage rather than as a data-generating mechanism. Both parametric models retained $A \times Z$ but omitted $A \times N$, although the latter interaction was present in the data generating mechanism. This simpler specification is relevant in applications involving many care settings and multiple non-allowable covariates. Allowing the coefficient of each non-allowable covariate to vary across settings can substantially increase the number of parameters. These coefficients can be difficult to estimate when some settings have limited data. By including \(A\times N\) in the data generating mechanism and omitting it from the fitted outcome models, we assess estimator performance under this form of outcome-model misspecification.

As a comparison to the parametric outcome models, we also used machine learning algorithms, namely random forest and XGBoost, to obtain predicted outcomes for the variance component estimators. Within each replicate, we used grid search with five-fold cross-validation for hyperparameter tuning. For random forest, we tuned the number of variables randomly sampled at each split (\textit{mtry}) and the minimum node size, using Gini splitting for classification and variance splitting for regression \citep{probst2019hyperparameters}. For XGBoost, the tuning grid included the number of boosting iterations, maximum tree depth, shrinkage (\textit{eta}), minimum loss reduction, subsample ratio of columns, and subsample percentage. Details of the hyperparameter tuning process and search grids are provided in Appendix~\ref{sec:AH_hyperparam_tune}. XGBoost used logistic loss for dichotomous outcomes and squared-error loss for continuous outcomes; random forest used the corresponding impurity-based splitting rules. Hyperparameters were selected by maximizing the mean validation area under the ROC curve (AUC) across folds for dichotomous outcomes and minimizing the mean validation root mean squared error (RMSE) across folds for continuous outcomes.

We also evaluated the performance of the two uncertainty propagation procedures described in Section~\ref{sec:identification-estimation}. Given the computational cost, we focused this assessment on the fixed-effects estimators with a binary outcome as an illustration. Each procedure used 1,000 draws per simulation replicate. Within each replicate, the standard deviation across draws was used as the estimated standard error for each component. For each scenario, we compared the average of these standard errors with the Monte Carlo standard deviation of the corresponding plug-in point estimates. Similar simulation-based assessments could be conducted for other continuous outcomes and model specifications.

\subsection{Simulation Results}
We present our simulation results for a dichotomous outcome. The same simulation was repeated for a continuous outcome, with the results provided in Appendix~\ref{sec:AJ_cont_res}. Figure~\ref{fig:hosp5_binary_res} presents the sampling distributions of point estimates for each component across 1,000  replicates
when \(J=5\), under varying sample sizes. Each panel corresponds to one variance component. Estimators based on different outcome models are distinguished by fill shading. The true values are shown as horizontal dashed lines.

The fixed-effects estimates exhibited some small-sample bias, particularly a positive one for the effect modification component \(V_{\mathrm{em}}\), which diminished as sample size increased. The random-effects estimates on the other hand showed a negative bias which shrunk with increasing sample size. Overall, the mixed-effects model produced less variable estimates, which is expected due to shrinkage. Estimators based on nonparametric outcome models did not consistently approach the true values as sample size increased; some components were systematically underfitted and some overfitted, but overall the result was underfit, reflected by overestimation of the residual variance component.

For estimators based on parametric outcome models, although the fitted fixed- and mixed-effects outcome models omitted \(A\times\boldsymbol N\), their estimates were close to the true values for most components at larger sample sizes in the simulation scenarios considered. The decomposition uses outcome means averaged over the natural distribution of \(\boldsymbol N\) within each \((Z,\boldsymbol X)\) stratum. Component estimates may therefore remain accurate when these averaged means are well approximated, even if the conditional outcome model omits \(A\times\boldsymbol N\). However, these findings do not establish general robustness to outcome-model misspecification.

\begin{figure}
    \centering
    \includegraphics[width=1\linewidth]{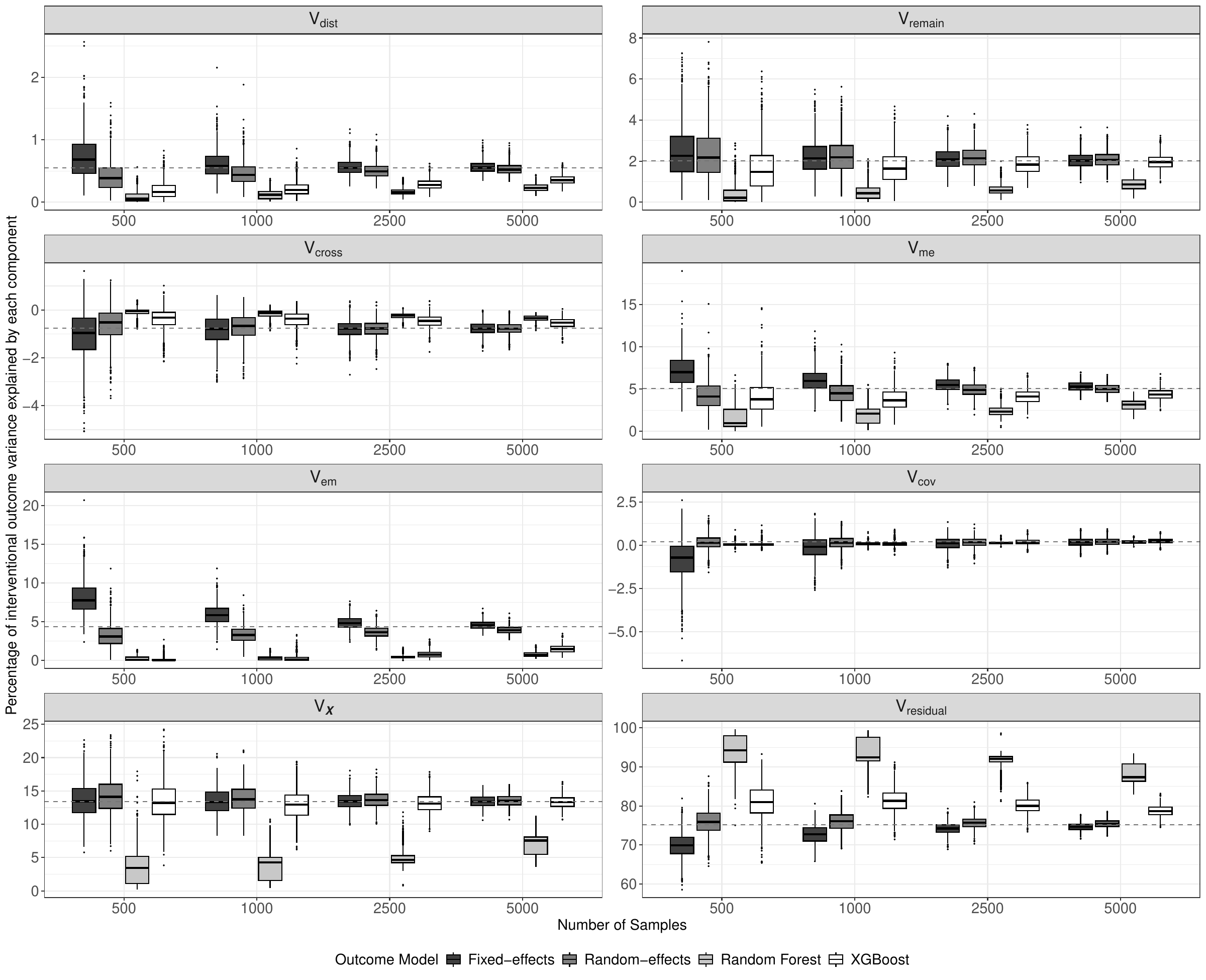}
    \caption{Sampling distributions for the estimated percentage of
    total interventional outcome variance contributed
    by each component, under different sample sizes across
    1,000  replicates, with the number of
    settings \(J=10\). The horizontal dashed lines
    indicate the true values.
    FE and RE outcome models omit \(A\times N\) interaction.}
    \label{fig:hosp10_binary_res}
\end{figure}

When \(J=10\) (Figure~\ref{fig:hosp10_binary_res}), the patterns were broadly similar to those observed with \(J=5\).
Small-sample bias in the fixed-effects estimates of \(V_{\mathrm{em}}\) became more apparent, while the random-effects estimates remained below the benchmark. It is not surprising that the effect modification estimates are sensitive to the number of settings, as sufficient data in each cross-classification category is needed to estimate the corresponding interaction terms. Increasing sample size reduced the small-sample discrepancies. The ML estimates also continued to underestimate effect heterogeneity, and underfit overall. The ML results may reflect regularization bias, when the models are tuned to predict out of sample, rather than targeted towards the specific quantities of interest.

\begin{figure}
    \centering
    \includegraphics[width=1\linewidth]{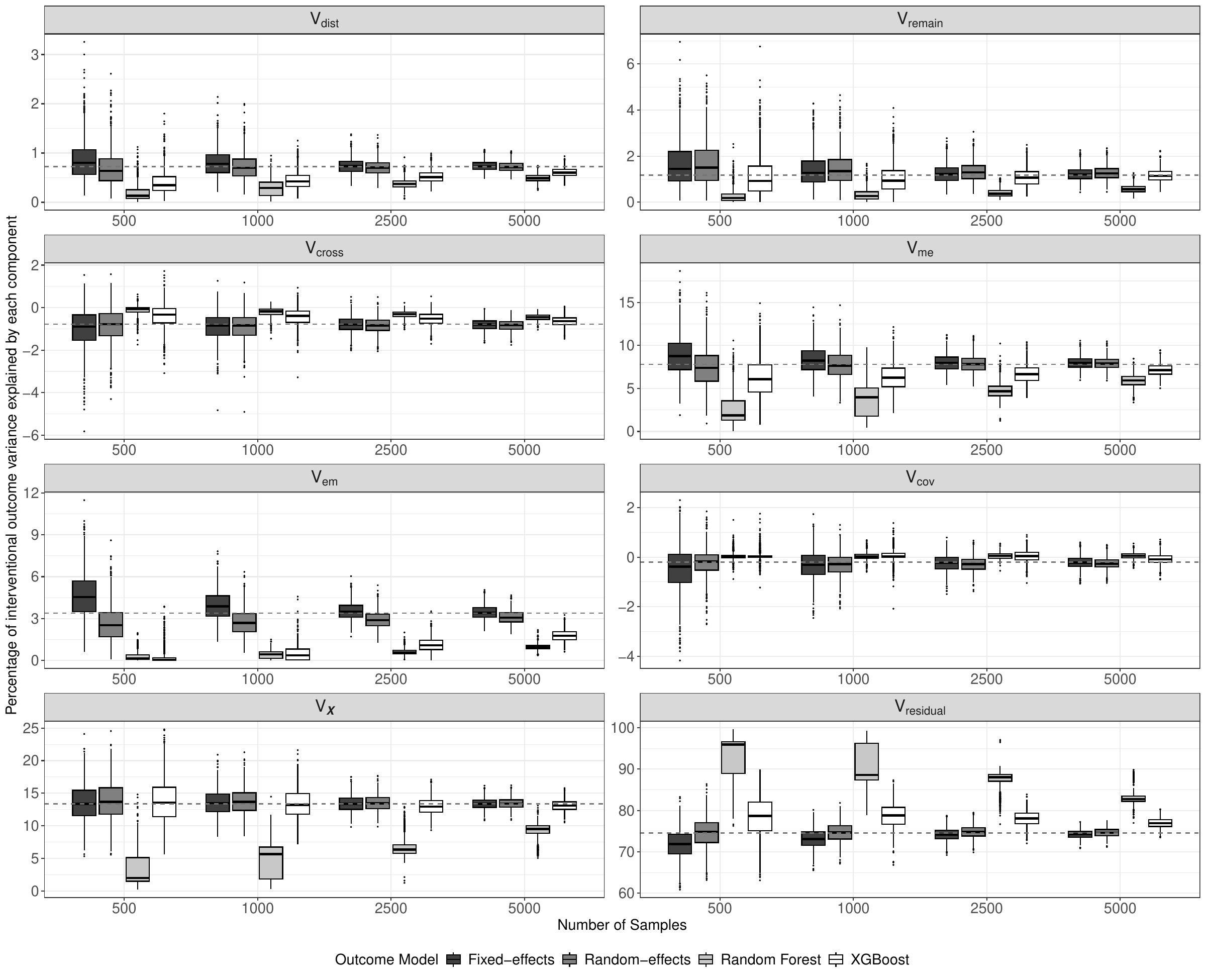}
    \caption{Sampling distributions for the estimated percentage of
   total interventional outcome variance contributed
    by each component, under different sample sizes across 1,000  replicates, with the number of
    settings \(J=5\). The horizontal dashed lines indicate the true values.
    FE and RE outcome models omit \(A\times N\) interaction.}
    \label{fig:hosp5_binary_res}
\end{figure}
\FloatBarrier

\FloatBarrier

\begin{figure*}[!htbp]
\centering
\includegraphics[width=\textwidth]
{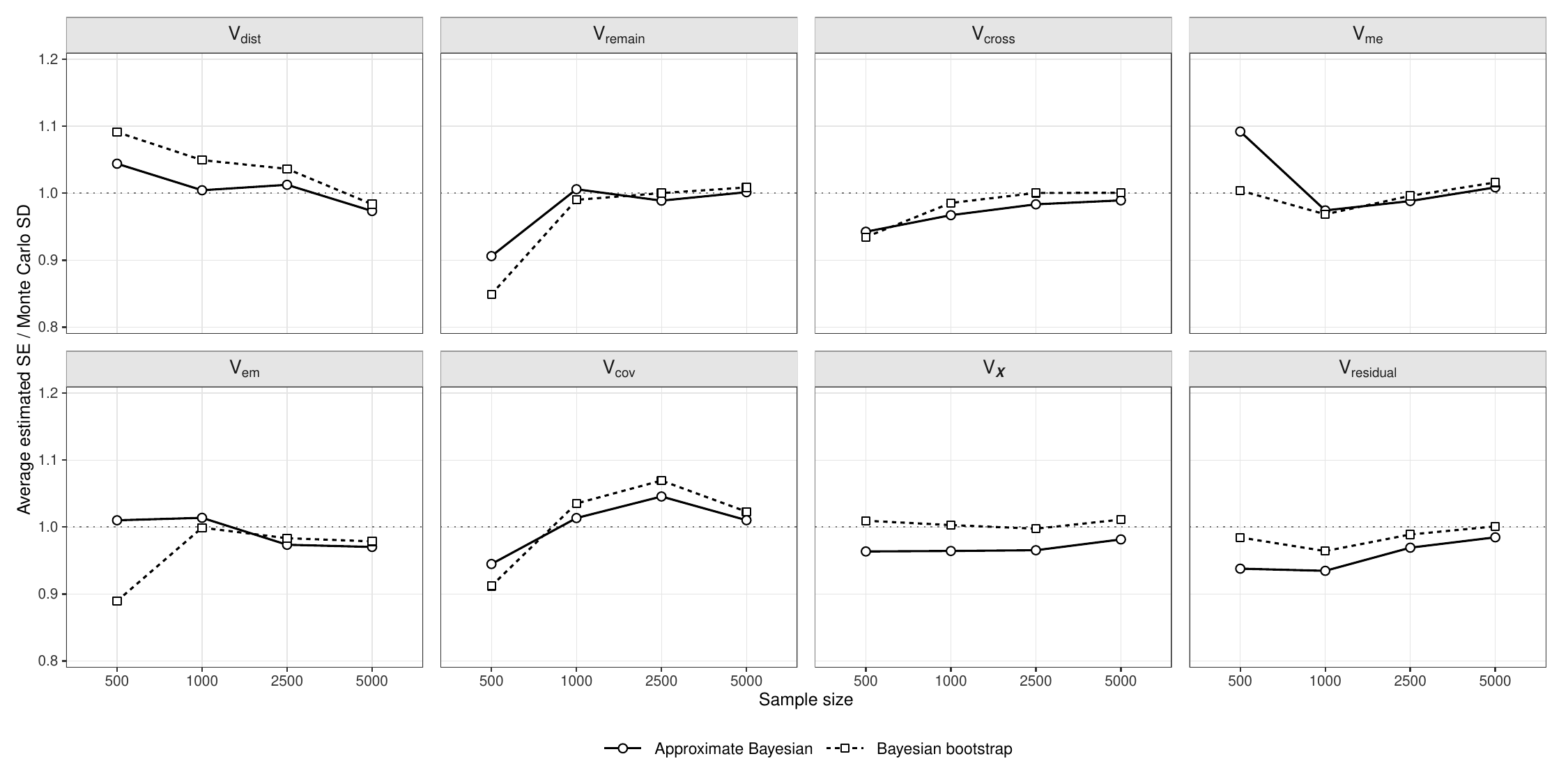}
\caption{Ratio of the average estimated SE to the Monte Carlo SD
for each variance component of a binary outcome, using fixed-effects
estimators with \(J=10\) care settings.}
\label{fig:SE_MCSD_ratio_J10}
\end{figure*}

Figure~\ref{fig:SE_MCSD_ratio_J10} compares the average estimated standard errors (SEs) from approximate Bayesian parameter sampling and the Bayesian bootstrap with the Monte Carlo standard deviations (SDs) of the fixed-effects estimators for a binary outcome, with \(J=10\) care settings. Results are presented as ratios of average SE to Monte Carlo SD, with 1 indicating agreement. Each panel corresponds to one variance component, evaluated on its original variance scale. Solid lines with open circles denote approximate Bayesian parameter sampling, and dashed lines with open squares denote the Bayesian bootstrap.

The results indicated that average SEs from both procedures were generally close to the Monte Carlo SDs, with closer agreement at larger sample sizes. Discrepancies were more pronounced at smaller sample sizes. Neither procedure consistently showed closer agreement to Monte Carlo SDs across all components. For example, with \(J=10\) and \(n=500\), approximate Bayesian sampling yielded SEs closer to the Monte Carlo SD for  \(V_{\mathrm{em}}\), whereas the Bayesian bootstrap was closer for \(V_{\mathrm{me}}\). The corresponding results for \(J=5\) showed a similar overall pattern (see Appendix~\ref{sec:SE_MCSD_ratio_J5}).

\FloatBarrier

\section{Illustration Using the SEER Database}\label{sec:seer}
We illustrate the proposed decomposition using data from the Surveillance, Epidemiology, and End Results (SEER) Program \citep{seerHome2025} to examine geographic variation in brachytherapy receipt and its relationship with racial and ethnic group membership among women diagnosed with cervical cancer. Brachytherapy is an essential component of definitive chemoradiation for locally advanced cervical cancer. However, a SEER analysis of 8,500 cervical cancer patients found that brachytherapy use varied by geographic region and demographic and socioeconomic characteristics \citep{han2024updated}. These findings motivate examining how geographic context, differences in registry-area distributions across racial and ethnic groups, and group differences in outcomes between geographic areas contribute to outcome variation. These geographic comparisons can guide further investigation of regional referral pathways, treatment capacity, and the availability of brachytherapy services.

Our analytic sample included 8,072 patients from the SEER cohort studied by Han et al.~\citep{han2024updated}. The process-of-care outcome \(Y\) was a binary indicator of brachytherapy receipt, compared to the alternative External Beam Radiation Therapy (EBRT). The care setting variable \(A\) denotes the SEER registry area corresponding to each patient's residence at diagnosis. We interpret the resulting contrasts as residence-based geographic comparisons of brachytherapy receipt. The 11 registry areas in the analysis were California, Connecticut, Georgia, Hawaii, Iowa, Kentucky, Louisiana, New Jersey, New Mexico, Seattle (Puget Sound), and Utah.

Group membership $Z$ comprised four mutually exclusive categories: non-Hispanic Black, Hispanic (all races), non-Hispanic Asian or Pacific Islander, and non-Hispanic White. These categories were constructed using the SEER race and origin recode \cite{seerRaceOriginRecode}. Patients identified as Hispanic were classified as Hispanic regardless of recorded race, whereas the other categories were restricted to non-Hispanic patients. Non-Hispanic patients with unknown race were excluded. Non-Hispanic American Indian/Alaska Native patients were also excluded because sparse counts across registry areas did not support reliable estimation as a separate group. Appendix~\ref{sec:seer_eda} describes the exclusion criteria and summarizes the characteristics of the analytic sample.

We incorporate covariates identified by Han et al. \cite{han2024updated} into our analysis. We classified age at diagnosis, cancer stage, and year of diagnosis as fully allowable covariates \(\boldsymbol X\). Age and stage are relevant to clinical decisions about brachytherapy, and conditioning on year of diagnosis allows us to compare patients diagnosed during the same period of clinical practice. We therefore define the disparity within these covariate profiles and allow the hypothetical assignment distribution to vary across clinical profiles and calendar years. We classified metro residence, marital status, and income category as fully non-allowable covariates \(\boldsymbol N\) because they reflect geographic and socioeconomic circumstances that may shape access to brachytherapy. Therefore, we considered differences in brachytherapy receipt associated with these circumstances part of the disparity of interest.

To assess data support and model fitting, we examined registry-by-group sample sizes and outcome counts, covariate overlap, model convergence, and fitted assignment probabilities. We fitted two multinomial logistic regression models, one for registry assignment, including racial and ethnic group membership, allowable covariates $\boldsymbol X$, and non-allowable covariates $\boldsymbol N$ as predictors; and another for group membership, using only $\boldsymbol X$ as predictors. To address identifiability issues arising from small covariate categories in some registries, we constrained selected coefficients to zero. Rather than defining a fixed threshold for small cell counts, we imposed the minimal necessary constraints to ensure model identifiability \citep{tang2020constructing}. We modeled the joint distribution $f(\boldsymbol N\mid Z,\boldsymbol X)$ sequentially, in the order metro residence, marital status, and income. Binary logistic regression was used for metro residence and marital status, and proportional-odds ordinal logistic regression was used for the three income categories. Each model included group membership, $\boldsymbol X$, and the preceding non-allowable covariates.

We specified and fitted four outcome models: (i) a logistic fixed-effects model including registry, group membership, their interaction, and both $\boldsymbol X$ and $\boldsymbol N$; (ii) a logistic mixed-effects model with group membership, $\boldsymbol X$, and $\boldsymbol N$ as fixed effects, a random intercept for registry, and random slopes for group contrasts within each registry; and (iii) random forest (RF) and (iv) XGBoost models using registry, group membership, $\boldsymbol X$, and $\boldsymbol N$ as inputs. Hyperparameters for the RF and XGBoost models were tuned via five-fold cross-validation. The tuning procedure and search grids are described in Appendix~\ref{sec:AH_hyperparam_tune}. To assess estimator variability, we employed 1,000 Bayesian bootstrap draws, following the procedure described in Section \ref{sec:identification-estimation}. For comparison, we also quantified uncertainty for the fixed-effects estimates using the approximate Bayesian parameter sampling procedure described in the same section.

The variance decomposition results based on the Bayesian bootstrap are presented in Table ~\ref{tab:seer_interventional_percent}. Columns correspond to different outcome models, and rows report proportional variance components, expressed as percentages of the total interventional outcome variance, with bootstrap medians reported as point estimates and 95\% credible intervals shown in brackets. Table ~\ref{tab:seer_interventional_values} reports the corresponding components on their original scale (multiplied by 100 for presentation), and bootstrap sampling distributions are provided in Appendix \ref{sec:AL_seer_sample_dist}. For the fixed-effects model, we also computed 95\% credible intervals using approximate Bayesian parameter sampling. These were generally wider than those obtained using the Bayesian bootstrap.

Total interventional outcome variance was similar across the four outcome models, with Bayesian bootstrap medians ranging from 0.2298 to 0.2312. The residual component accounted for 91.87\%–94.13\% of this variance. The residual component captures outcome variation remaining within assignment, group, and allowable covariate strata under the stochastic intervention. For a binary outcome, substantial within-stratum variation can remain even when the conditional outcome probabilities are correctly specified. Therefore, a large residual share does not indicate model inadequacy or invalidate the decomposition. Also, the prominence of the residual component does not indicate limited usefulness of the decomposition. Rather, the goal of the approach is not to maximize explained variance, but to assess contributions associated with prespecified interventional contrasts on a common variance scale. Among the remaining components, the main care-setting component was the largest, accounting for 3.25\%–4.79\% of the total variance and summarizing average outcome contrasts between registry areas. Overall, the group components (\(V_{\mathrm{dist}}, V_{\mathrm{remain}}, V_{\mathrm{cross}}\)), the effect-modification component (\(V_{\mathrm{em}}\)), and the assignment--effect covariance component (\(V_{\mathrm{cov}}\)) were relatively small in magnitude compared with total interventional outcome variance in brachytherapy receipt. The care-setting distribution (\(V_{\mathrm{dist}}\)) and remaining group (\(V_{\mathrm{remain}}\)) components accounted for 0.11\%–0.20\% and 0.09\%–0.28\%, respectively. The cross component \(V_{\mathrm{cross}}\) ranged from \(-0.03\%\) to \(0.13\%\). Covariance component \(V_{\mathrm{cov}}\) had negative medians under the fixed- and mixed-effects models (\(-1.19\%\) and \(-0.26\%\), respectively) and medians close to zero under RF and XGBoost.

The allocation of variance across some of the components was sensitive to outcome model specification. In particular, the effect modification component accounted for 1.73\% under the fixed-effects model, 0.67\% under the mixed-effects model, 0.06\% under random forest, and 0.02\% under XGBoost. The allowable covariate component was larger under the nonparametric models (RF and XGBoost). This difference may reflect their flexibility in capturing covariate interactions that were not explicitly included in the parametric specifications.

\FloatBarrier
\input{Tables/table1_percent_BB}

\FloatBarrier
\input{Tables/table2_values_x100_BB}
\FloatBarrier

\section{Discussion}\label{sec:discussion}
In this paper, we developed a causal variance decomposition framework that decomposes outcome variance under a prespecified stochastic assignment intervention and incorporates explicit allowability designations to define the disparity of interest and the intervention. It separates group-related variation into components associated with care-setting distributions, remaining group contrasts, and their cross component. It also distinguishes average variation between care settings, heterogeneity in setting contrasts across groups, and assignment--effect covariance. These components summarize variation across multiple settings and groups on a common variance scale, providing a first step toward identifying potential targets for intervention in treatment delivery and service provision.

The conditioning order in our decomposition is guided by the disparity of interest. Following the allowability framework \citep{jackson2021meaningful,chang2024importance}, group comparisons are defined within fully allowable covariate profiles, while differences in the natural conditional distribution of the non-allowable covariates \(\boldsymbol N\) are retained. We therefore condition first on \(\boldsymbol X\) and then on group membership, so that the group component summarizes variation in expected interventional outcomes across groups with the same allowable covariate profile. Non-allowable covariates are not introduced as a separate conditioning stage in the decomposition but are still included in nuisance models for confounding control. The split of relevant confounders/intermediate variables into  \(\boldsymbol X\) and  \(\boldsymbol N\) subsets depends on the context and desired disparity contrast. Our choice of conditioning is also conceptually related to the notion of ``counterfactual fairness'' considered by Karvanen et al. \cite{karvanen2024simulating}.

As we conceptualized the case-mix covariates to be on the pathway downstream from the group indicator, as is common in disparity contexts \citep{naimi2016mediation}, our conditioning order then does not follow the topological order of the DAG characterizing the data generating mechanism \citep{saarela2026graph}. As a result, while the case-mix conditional terms are interpretable, the case-mix variance term in the decomposition does not in itself have a causal interpretation, which is not a major limitation as we could have limited ourselves to decomposing the case-mix conditional variance (i.e., \(E_{\boldsymbol X}\{V(Y^{\mathcal G}\mid\boldsymbol X)\}\)).  Alternative decompositions are possible, depending on the estimands of interest, with different resulting interpretations. For example, conditioning jointly on \(\boldsymbol X\) and \(\boldsymbol N\) (or combining \(\boldsymbol N\) covariates with \(\boldsymbol X\)) before group membership would define group comparisons among individuals with the same values of both covariate sets. This resembles conventional hospital profiling, where case-mix adjustment facilitates comparisons of care among patients with similar measured characteristics \citep{hartman2024evaluating,valeri2023multistate}. Another alternative would place group membership before the allowable covariates \(\boldsymbol X\), following their topological order more closely. The resulting group component would compare expected outcomes averaged over each group's natural distribution of \(\boldsymbol X\), thereby also reflecting group differences in \(\boldsymbol X\) and associated setting assignment patterns.

The current framework adopts a two-set allowability designation, distinguishing fully allowable from fully non-allowable covariates for both disparity definition and intervention specification. We consider this formulation appropriate for the process-of-care outcomes of the type examined here, where pre-treatment clinical characteristics in \(\boldsymbol X\) can justify differences in treatment receipt and are also permitted to guide assignment under the intervention. However, a covariate’s allowability designation may differ between disparity definition and intervention specification, depending on the setting. For example, cancer stage at diagnosis may be considered outcome-allowable when comparing a process-type outcome, such as treatment receipt, because it helps define clinically appropriate care. When studying an outcome-type measure such as complications or survival, however, stage differences may be retained as part of the disparity if the research question includes inequalities associated with delayed diagnosis. Other applications may therefore benefit from accommodating mixed allowability designations. We are currently extending the framework to accommodate separate sets of outcome-allowable and treatment-allowable covariates and will address this in a separate work.

For estimation, we proposed model-based plug-in estimators using outcome, care-setting assignment, group-membership, and non-allowable covariate models. These models correspond to a factorization of the observed-data likelihood, which facilitates Bayesian or approximate Bayesian inference on their parameters and produces joint inferences for the full variance decomposition simultaneously. Recomputing all components from each joint parameter draw propagates uncertainty through the decomposition while preserving the components' dependence and additive relationships. However, misspecification of any of these models may affect the plug-in estimates. Flexible machine-learning models, as included in our simulations as comparisons, provide alternatives to parametric specifications, but regularization/hyperparameter tuning optimized for out-of-sample prediction may introduce bias when the fitted values are used directly in plug-in estimators. Efficient influence function-based approaches offer strategies for reducing sensitivity to nuisance-estimation errors under suitable conditions \citep{williamson2021nonparametric,williamson2023general,yu2025nonparametric}. Although addressing regularization bias is an important direction and will be developed as future work, developing or applying correction methods/targeted estimators is beyond the scope of the present paper, which focused on new disparity-relevant variance components and their causal interpretation.

Additional estimation challenges arise when the analysis includes many care settings and some settings have limited data. In the fixed-effects outcome model, setting-by-group interactions may be difficult to estimate when some combinations contain few patients or, for binary outcomes, few events. Partial pooling in mixed-effects models can stabilize these estimates, with shrinkage potentially attenuating differences across settings and their heterogeneity across groups \citep{varewyck2014shrinkage}. The care-setting assignment model presents a related challenge. A conventional multinomial logistic model estimates a separate coefficient vector for each nonreference setting, which can make estimation unstable when some settings have limited data \citep{tang2020constructing,varewyck2014shrinkage}. One option is to use a constrained specification that fixes selected coefficients at zero for settings with sparse data. Another option is to fit separate binary logistic regressions, each comparing one non-reference setting with a common reference setting, and then normalize the fitted relative odds to obtain assignment probabilities across all settings \citep{begg1984calculation}. When covariate overlap is limited, some setting comparisons may rely heavily on model extrapolation even when the assignment model converges. There is no universal cutoff for the number of care settings that can be accommodated. To assess feasibility in a given application, we recommend examining setting-specific sample sizes, group representation, and outcome counts, along with covariate overlap, model convergence, and estimated assignment probabilities.

We are currently working on several extensions of this work. One possible direction is to extend the framework to time-to-event outcomes, potentially through the use of restricted mean survival time \cite{han2022restricted}. Another potential extension is to consider a sequential or mediation-type framework in which process-type quality indicators (e.g., treatment or procedure received by patients) are modeled as mediators linking hospital characteristics to outcome-type quality indicators (e.g., survival), similar to Chen et al. \cite{chen2022causal}. In this setting we could potentially study moderated mediation \cite{muller2005moderation} by the sociodemographic group membership.

\section*{Acknowledgments}

For the original submission, the authors wrote the manuscript and developed the variance decomposition and analysis code for the case in which all measured covariates were designated as allowable. They used a chatbot (Microsoft 365 Copilot Chat using the GPT-5.4 model) to check spelling, grammar, consistency of mathematical notation, and mathematical derivations.

During revision, the authors developed the interventional extension including non-allowable covariates with feedback from chatbots (ChatGPT and Microsoft 365 Copilot Chat using GPT-5.5–5.6 models). They used a coding agent (OpenAI Codex using GPT-5.6–6.2 models) to help refine mathematical formulations and check derivations in the manuscript and supplementary appendices, draft and edit passages for clarity and readability, manage the bibliography, and check the responses to reviewers against the revised manuscript. They also used Codex to check, modify, and extend their existing R code for simulation, statistical analysis, numerical diagnostics, and visualization. The authors originated the scientific ideas, directed and reviewed the AI-assisted work, and take full responsibility for the scientific content, code, analyses, and final interpretations.

OpenAI provided complimentary access to its academic plan used during the revision.

\section*{Financial disclosure}

The work of the last author was supported by Discovery Grants RGPIN-2020-05920 and RGPIN-2026-06817 from the Natural Sciences and Engineering Research Council of Canada.

\section*{Conflict of interest}

The authors declare no potential conflicts of interest.

\clearpage
\bibliographystyle{unsrtnat}
\providecommand{\doi}[1]{\href{https://doi.org/#1}{doi:\nolinkurl{#1}}}
\bibliography{bibliography}

\clearpage
\section*{Supporting information}
\input{supplementary}
\end{document}

%% file: Figures/DAG_include_U.tex
\begingroup

\newcommand{\selectionpanel}[2]{%
  \begin{tikzpicture}[
    x=1cm,y=1cm,
    >={Stealth[length=2.4mm,width=1.6mm]},
    dag node/.style={circle,draw=black!70,fill=white,
                     minimum size=7mm,inner sep=0pt},
    observed edge/.style={->,semithick,draw=black!75,
                          shorten >=1pt,shorten <=1pt},
    latent edge/.style={observed edge,dashed,draw=black}
  ]
    \path[use as bounding box] (-1.2,-2.55) rectangle (6.5,3.65);
    \node[font=\large] at (2.75,3.35) {#1};

    \node[dag node] (Z) at (0,0) {$Z$};
    \node[dag node] (N) at (2,1.45) {$\boldsymbol N$};
    \node[dag node] (X) at (2,-1.45) {$\boldsymbol X$};
    \node[dag node] (A) at (4,0) {$A$};
    \node[dag node] (Y) at (6,0) {$Y$};

    \draw[observed edge] (Z) -- (N);
    \draw[observed edge] (Z) -- (X);
    \draw[observed edge] (Z) -- (A);
    \draw[observed edge] (Z.south)
      .. controls (0,-2.95) and (6,-2.95) .. (Y.south);

    \draw[observed edge] (N) -- (X);
    \draw[observed edge] (N) -- (A);
    \draw[observed edge] (N.east)
      .. controls (3.65,1.52) and (5.10,1.24) .. (Y.north west);

    \draw[observed edge] (X) -- (A);
    \draw[observed edge] (X.east)
      .. controls (3.65,-1.52) and (5.10,-1.24) .. (Y.south west);
    \draw[observed edge] (A) -- (Y);

    #2
  \end{tikzpicture}%
}

\newcommand{\assignmentselection}{%
  \node[dag node] (Uone) at (0.25,2.58) {$U_1$};
  \draw[latent edge] (Uone) -- (N);
  \draw[latent edge] (Uone.west)
    .. controls (-1.7,1.9) and (-1.25,-1.30) .. (X.south west);
  \draw[latent edge] (Uone.east)
    .. controls (2.50,2.68) and (3.82,1.68) .. (A.north);
}

\newcommand{\outcomeselection}{%
  \node[dag node] (Utwo) at (5.95,2.58) {$U_2$};
  \draw[latent edge] (Utwo) -- (N);
  \draw[latent edge] (Utwo.south west)
    .. controls (3.12,1.90) and (2.77,-0.20) .. (X.65);
  \draw[latent edge] (Utwo) -- (Y);
}

\begin{figure}[!htbp]
  \centering
  \resizebox{0.32\linewidth}{!}{%
    \selectionpanel{(a) Baseline graph}{}%
  }\hfill
  \resizebox{0.32\linewidth}{!}{%
    \selectionpanel{(b) Assignment selection}{\assignmentselection}%
  }\hfill
  \resizebox{0.32\linewidth}{!}{%
    \selectionpanel{(c) Outcome selection}{\outcomeselection}%
  }

\caption{Directed acyclic graphs showing the assumed causal
relationships among group membership $Z$, allowable covariates $\boldsymbol{X}$, non-allowable covariates $\boldsymbol{N}$, care-setting assignment $A$, and the outcome $Y$. Panel (a) shows the baseline graph with only observed variables. Panels (b) and (c) introduce assignment-related selection through unmeasured variable \(U_1\) and outcome-related selection through unmeasured variable \(U_2\), respectively. Dashed arrows denote causal relationships involving unmeasured characteristics.}
  \label{fig:DAG_include_U}
\end{figure}

\endgroup

%% file: Tables/table1_percent_BB.tex
\begin{table}[!htbp]
\centering
\caption{Interventional variance decomposition as percentages of total interventional variance across outcome models.}
\label{tab:seer_interventional_percent}
\begingroup
\setlength{\tabcolsep}{4pt}
\begin{tabularx}{\linewidth}{@{}>{\raggedright\arraybackslash}p{0.16\linewidth}*{4}{>{\centering\arraybackslash}X}@{}}
\toprule
\textbf{Component} & \textbf{FE} & \textbf{RE} & \textbf{RF} & \textbf{XGBoost} \\
\midrule
\(V_{\mathrm{dist}}\) & 0.20\% & 0.17\% & 0.11\% & 0.13\% \\
 & [0.09\%, 0.35\%] & [0.08\%, 0.30\%] & [0.06\%, 0.18\%] & [0.08\%, 0.20\%] \\
\(V_{\mathrm{remain}}\) & 0.28\% & 0.24\% & 0.11\% & 0.09\% \\
 & [0.08\%, 0.62\%] & [0.06\%, 0.58\%] & [0.00\%, 0.25\%] & [0.00\%, 0.32\%] \\
\(V_{\mathrm{cross}}\) & -0.03\% & 0.02\% & 0.13\% & 0.07\% \\
 & [-0.34\%, 0.22\%] & [-0.26\%, 0.24\%] & [0.02\%, 0.25\%] & [-0.07\%, 0.22\%] \\
\(V_{\mathrm{me}}\) & 4.79\% & 4.06\% & 3.25\% & 3.92\% \\
 & [3.63\%, 6.09\%] & [2.87\%, 5.29\%] & [2.11\%, 4.45\%] & [2.69\%, 5.27\%] \\
\(V_{\mathrm{em}}\) & 1.73\% & 0.67\% & 0.06\% & 0.02\% \\
 & [1.00\%, 2.75\%] & [0.23\%, 1.45\%] & [0.00\%, 0.18\%] & [0.00\%, 0.21\%] \\
\(V_{\mathrm{cov}}\) & -1.19\% & -0.26\% & 0.03\% & -0.01\% \\
 & [-2.05\%, -0.32\%] & [-1.10\%, 0.45\%] & [-0.11\%, 0.20\%] & [-0.24\%, 0.10\%] \\
\(V_{\boldsymbol X}\) & 0.88\% & 0.92\% & 2.57\% & 3.88\% \\
 & [0.52\%, 1.36\%] & [0.50\%, 1.49\%] & [1.36\%, 3.94\%] & [2.60\%, 5.25\%] \\
\(V_{\mathrm{residual}}\) & 93.26\% & 94.13\% & 93.69\% & 91.87\% \\
 & [92.10\%, 94.37\%] & [93.00\%, 95.23\%] & [91.06\%, 95.95\%] & [89.24\%, 94.41\%] \\
\bottomrule
\end{tabularx}
\endgroup
\par\smallskip
\begin{minipage}{\linewidth}\footnotesize
\end{minipage}
\end{table}

%% file: Tables/table2_values_x100_BB.tex
\begin{table}[!htbp]
\centering
\caption{Interventional variance decomposition in original units (multiplied by 100 for presentation)  across outcome models.}
\label{tab:seer_interventional_values}
\begingroup
\setlength{\tabcolsep}{4pt}
\begin{tabularx}{\linewidth}{@{}>{\raggedright\arraybackslash}p{0.16\linewidth}*{4}{>{\centering\arraybackslash}X}@{}}
\toprule
\textbf{Component} & \textbf{FE} & \textbf{RE} & \textbf{RF} & \textbf{XGBoost} \\
\midrule
\(V(Y^{\mathcal G})\) & 23.11 & 23.12 & 22.98 & 23.06 \\
 & [22.78, 23.42] & [22.81, 23.40] & [22.69, 23.27] & [22.75, 23.36] \\
\(V_{\mathrm{dist}}\) & 0.05 & 0.04 & 0.03 & 0.03 \\
 & [0.02, 0.08] & [0.02, 0.07] & [0.01, 0.04] & [0.02, 0.05] \\
\(V_{\mathrm{remain}}\) & 0.06 & 0.06 & 0.03 & 0.02 \\
 & [0.02, 0.14] & [0.01, 0.14] & [0.00, 0.06] & [0.00, 0.07] \\
\(V_{\mathrm{cross}}\) & -0.01 & 0.01 & 0.03 & 0.02 \\
 & [-0.08, 0.05] & [-0.06, 0.05] & [0.00, 0.06] & [-0.02, 0.05] \\
\(V_{\mathrm{me}}\) & 1.11 & 0.94 & 0.75 & 0.90 \\
 & [0.84, 1.41] & [0.66, 1.23] & [0.48, 1.02] & [0.62, 1.21] \\
\(V_{\mathrm{em}}\) & 0.40 & 0.15 & 0.01 & 0.00 \\
 & [0.23, 0.63] & [0.05, 0.34] & [0.00, 0.04] & [0.00, 0.05] \\
\(V_{\mathrm{cov}}\) & -0.27 & -0.06 & 0.01 & 0.00 \\
 & [-0.47, -0.07] & [-0.25, 0.10] & [-0.03, 0.05] & [-0.05, 0.02] \\
\(V_{\boldsymbol X}\) & 0.20 & 0.21 & 0.59 & 0.89 \\
 & [0.12, 0.31] & [0.11, 0.35] & [0.32, 0.91] & [0.60, 1.21] \\
\(V_{\mathrm{residual}}\) & 21.55 & 21.75 & 21.52 & 21.19 \\
 & [21.18, 21.93] & [21.39, 22.11] & [20.94, 22.15] & [20.48, 21.83] \\
\bottomrule
\end{tabularx}
\endgroup
\par\smallskip
\begin{minipage}{\linewidth}\footnotesize
\end{minipage}
\end{table}

%% file: supplementary.tex
\appendix
\renewcommand{\theHequation}{appendix.\Alph{section}.\arabic{equation}}
\setcounter{equation}{0}

\makeatletter
\renewcommand{\@seccntformat}[1]{%
  \ifstrequal{#1}{section}{\appendixname\nobreakspace}{}%
  \csname the#1\endcsname\quad}
\makeatother
\renewcommand{\figurename}{Web Figure}
\renewcommand{\tablename}{Web Table}
\renewcommand{\theequation}{\Alph{section}.\arabic{equation}}
\counterwithin{figure}{section}
\counterwithin{table}{section}

\setcounter{equation}{0}

\section{Relationship Between Observed and Interventional
Variance Decompositions}
\label{app:observed-variance-N}

We first decompose the variance of the observed outcome \(Y\) by applying
the law of total variance successively, using a conditioning order
$\boldsymbol X \prec Z \prec A \prec Y$, which is purposefully chosen
different from the topological order of the variables in the DAG in Figure~\ref{fig:DAG_include_U}(a) of the main text \citep{saarela2026graph}.
Conditioning first on the allowable covariates \(\boldsymbol X\) gives
\[
V(Y)
=V_{\boldsymbol X}\!\left\{E(Y\mid\boldsymbol X)\right\}
+E_{\boldsymbol X}\!\left\{V(Y\mid\boldsymbol X)\right\}.
\]
Within each covariate profile, we decompose \(V(Y\mid\boldsymbol X)\)
by group \(Z\). Within each group and profile, we then decompose
\(V(Y\mid Z,\boldsymbol X)\) by care setting \(A\). Substituting these
two conditional decompositions gives
\begin{align*}
V(Y)
&=\underbrace{V_{\boldsymbol X}\!\left\{E(Y\mid\boldsymbol X)\right\}}
             _{V_{\boldsymbol X}^{\mathrm{obs}}}\\
&\quad+\underbrace{E_{\boldsymbol X}\!\left[
 V_{Z\mid\boldsymbol X}\!\left\{E(Y\mid Z,\boldsymbol X)\right\}\right]}
             _{V_{\mathrm{group}}^{\mathrm{obs}}}\\
&\quad+\underbrace{E_{\boldsymbol X,Z}\!\left[
 V_{A\mid Z,\boldsymbol X}\!\left\{E(Y\mid A,Z,\boldsymbol X)\right\}\right]}
             _{V_{\mathrm{setting}}^{\mathrm{obs}}}\\
&\quad+\underbrace{E_{\boldsymbol X,Z,A}\!\left\{V(Y\mid A,Z,\boldsymbol X)\right\}}
             _{V_{\mathrm{residual}}^{\mathrm{obs}}}.
\end{align*}
In this order, the group component compares mean outcomes across groups
at the same allowable covariate profile, with each group retaining its
observed distribution of care settings. The setting component compares
mean outcomes across settings within each group and profile.
Applying the same steps to \(Y^{\mathcal G}\), with the randomized setting
\(G\) in place of the observed setting \(A\), gives
equation~\eqref{eq:overall_interventional_decomposition} of the main text.

All components without an ``obs'' superscript refer to the
interventional decomposition in the main text.
Under the stated identification assumptions, define
\begin{align*}
D_N(z,\boldsymbol x)
&=E(Y\mid z,\boldsymbol x)
-E(Y^{\mathcal G}\mid z,\boldsymbol x)\\
&=\sum_a\int
E(Y\mid a,z,\boldsymbol n,\boldsymbol x)
\left\{
P(a\mid z,\boldsymbol n,\boldsymbol x)
-P(a\mid z,\boldsymbol x)
\right\}
\,dF(\boldsymbol n\mid z,\boldsymbol x),
\end{align*}
and
$$
\overline D_N(\boldsymbol x)
=\sum_z D_N(z,\boldsymbol x)P(z\mid\boldsymbol x).
$$
Consequently,
\begin{align*}
E(Y\mid z,\boldsymbol x)
&=E(Y^{\mathcal G}\mid z,\boldsymbol x)
+D_N(z,\boldsymbol x),\\
E(Y\mid\boldsymbol X=\boldsymbol x)
&=E(Y^{\mathcal G}\mid\boldsymbol X=\boldsymbol x)
+\overline D_N(\boldsymbol x).
\end{align*}

\paragraph*{Allowable-covariate component.}

The variance-of-a-sum identity gives
\begin{align*}
V_{\boldsymbol X}^{\mathrm{obs}}
&=V_{\boldsymbol X}
\left\{
E(Y^{\mathcal G}\mid\boldsymbol X)
+\overline D_N(\boldsymbol X)
\right\}\\
&=V_{\boldsymbol X}+\Delta_{\boldsymbol X},
\end{align*}
where
\begin{align*}
\Delta_{\boldsymbol X}
&=V_{\boldsymbol X}
\{\overline D_N(\boldsymbol X)\} +
2\operatorname{Cov}_{\boldsymbol X}
\left\{
E(Y^{\mathcal G}\mid\boldsymbol X),
\overline D_N(\boldsymbol X)
\right\}.
\end{align*}

\paragraph*{Group component.}

Similarly,
\begin{align*}
V_{\mathrm{group}}^{\mathrm{obs}}
&=E_{\boldsymbol X}\!\left[
V_{Z\mid\boldsymbol X}
\left\{
E(Y^{\mathcal G}\mid Z,\boldsymbol X)
+D_N(Z,\boldsymbol X)
\right\}
\right]\\
&=V_{\mathrm{group}}+\Delta_{\mathrm{group}}\\
&=V_{\mathrm{dist}}+V_{\mathrm{remain}}
+V_{\mathrm{cross}}+\Delta_{\mathrm{group}},
\end{align*}
where
\begin{align*}
\Delta_{\mathrm{group}}
&=E_{\boldsymbol X}\!\left[
V_{Z\mid\boldsymbol X}\{D_N(Z,\boldsymbol X)\}
\right]\\
&\quad+
2E_{\boldsymbol X}\!\left[
\operatorname{Cov}_{Z\mid\boldsymbol X}
\left\{
E(Y^{\mathcal G}\mid Z,\boldsymbol X),
D_N(Z,\boldsymbol X)
\right\}
\right].
\end{align*}

\paragraph*{Setting component.}

For each unordered pair $(a,a^*)$, define
\begin{align*}
\kappa_{aa^*}(z,\boldsymbol x)
&=E(Y\mid a,z,\boldsymbol x)
-E(Y\mid a^*,z,\boldsymbol x)
-\delta_{aa^*}(z,\boldsymbol x).
\end{align*}
Here, the observed setting means average over
$F(\boldsymbol N\mid A,Z,\boldsymbol X)$, whereas the
potential-outcome means in $\delta_{aa^*}$ average over
$F(\boldsymbol N\mid Z,\boldsymbol X)$.

Because the stochastic intervention preserves
$P(A\mid Z,\boldsymbol X)$, the observed and interventional
pairwise assignment weights are identical. Thus,
\begin{align*}
V_{\mathrm{setting}}^{\mathrm{obs}}
&=E_{\boldsymbol X,Z}\!\left\{
\sum_{a<a^*}\pi_{aa^*}(Z,\boldsymbol X)
\left[
\delta_{aa^*}(Z,\boldsymbol X)
+\kappa_{aa^*}(Z,\boldsymbol X)
\right]^2
\right\}\\
&=V_{\mathrm{setting}}+\Delta_{\mathrm{setting}}\\
&=V_{\mathrm{me}}+V_{\mathrm{em}}
+V_{\mathrm{cov}}+\Delta_{\mathrm{setting}},
\end{align*}
where
\begin{align*}
\Delta_{\mathrm{setting}}
&=E_{\boldsymbol X,Z}\!\left\{
\sum_{a<a^*}\pi_{aa^*}(Z,\boldsymbol X)
\kappa_{aa^*}(Z,\boldsymbol X)^2
\right\}\\
&\quad+
2E_{\boldsymbol X,Z}\!\left\{
\sum_{a<a^*}\pi_{aa^*}(Z,\boldsymbol X)
\delta_{aa^*}(Z,\boldsymbol X)
\kappa_{aa^*}(Z,\boldsymbol X)
\right\}.
\end{align*}

\paragraph*{Residual component.}

The residual components satisfy
$$
V_{\mathrm{residual}}^{\mathrm{obs}}
=
V_{\mathrm{residual}}+\Delta_{\mathrm{residual}},
$$
where
\begin{align*}
\Delta_{\mathrm{residual}}
=E_{\boldsymbol X,Z}\!\Bigg\{
\sum_a P(a\mid Z,\boldsymbol X)
\Big[
&V(Y\mid A=a,Z,\boldsymbol X)-
V\{Y(a)\mid Z,\boldsymbol X\}
\Big]
\Bigg\}.
\end{align*}

\paragraph*{Combined decomposition.}

Combining these identities yields
\begin{align*}
V(Y)
={}&V_{\boldsymbol X}
+V_{\mathrm{dist}}+V_{\mathrm{remain}}+V_{\mathrm{cross}}\\
&+V_{\mathrm{me}}+V_{\mathrm{em}}+V_{\mathrm{cov}}
+V_{\mathrm{residual}}\\
&+\Delta_{\boldsymbol X}
+\Delta_{\mathrm{group}}
+\Delta_{\mathrm{setting}}
+\Delta_{\mathrm{residual}}.
\end{align*}
The first eight terms sum to $V(Y^{\mathcal G})$. Therefore,
\begin{equation}
\boxed{
V(Y)
=
V(Y^{\mathcal G})
+\Delta_{\boldsymbol X}
+\Delta_{\mathrm{group}}
+\Delta_{\mathrm{setting}}
+\Delta_{\mathrm{residual}}.
}
\label{eq:observed-interventional-variance-bridge}
\end{equation}

The additional terms describe changes associated with replacing
the observed $\boldsymbol N$-dependent assignment mechanism
with the stochastic intervention.
They are not necessarily nonnegative, so removing assignment
dependence does not necessarily reduce total outcome variance.
If $A\perp\!\!\!\perp\boldsymbol N\mid(Z,\boldsymbol X)$,
the observed and interventional conditional moments coincide
under the identification assumptions, and all four adjustment
terms vanish.
In particular, if all covariates are allowable
($\boldsymbol N=\varnothing$), the observed and interventional
decompositions coincide component by component.
Even if assignment depends on $\boldsymbol N$, the same conclusion
holds if $Y(a)\perp\!\!\!\perp A\mid(Z,\boldsymbol X)$ for every setting $a$.
By consistency and this exchangeability condition, the first two moments
of $Y$ given $(A=a,Z,\boldsymbol X)$ equal those of $Y(a)$ given
$(Z,\boldsymbol X)$, so all four adjustment terms again vanish.

\section{Decomposition of $V_{\text{group}}$}
\label{sec:V_group_proof}

Adding and subtracting
\[
\sum_a E\{Y(a)\mid z,\boldsymbol x\}
\sum_{z^*} P(a\mid z^*,\boldsymbol x)P(z^*\mid\boldsymbol x)
\]
gives
\begin{align*}
\delta_{\mathrm{group}}(z,\boldsymbol x)
&=
\sum_a E\{Y(a)\mid z,\boldsymbol x\}P(a\mid z,\boldsymbol x)\\
&\quad-
\sum_{z^*,a} E\{Y(a)\mid z^*,\boldsymbol x\}
P(a\mid z^*,\boldsymbol x)P(z^*\mid\boldsymbol x)\\
&=
\underbrace{
\sum_a E\{Y(a)\mid z,\boldsymbol x\}
\left\{
P(a\mid z,\boldsymbol x)
-\sum_{z^*} P(a\mid z^*,\boldsymbol x)P(z^*\mid\boldsymbol x)
\right\}
}_{\delta_{\mathrm{dist}}(z,\boldsymbol x)}\\
&\quad+
\underbrace{
\begin{aligned}
&\sum_a E\{Y(a)\mid z,\boldsymbol x\}
\sum_{z^*} P(a\mid z^*,\boldsymbol x)P(z^*\mid\boldsymbol x)\\
&\quad-
\sum_{z^*,a} E\{Y(a)\mid z^*,\boldsymbol x\}
P(a\mid z^*,\boldsymbol x)P(z^*\mid\boldsymbol x)
\end{aligned}
}_{\delta_{\mathrm{remain}}(z,\boldsymbol x)}.
\end{align*}
Hence,
\begin{align*}
V_{\mathrm{group}}
&=
E_{\boldsymbol X}\!\left\{
\sum_z
\left[
\delta_{\mathrm{dist}}(z,\boldsymbol X)
+\delta_{\mathrm{remain}}(z,\boldsymbol X)
\right]^2P(z\mid\boldsymbol X)
\right\}\\
&=
\underbrace{
E_{\boldsymbol X}\!\left\{
\sum_z\delta_{\mathrm{dist}}(z,\boldsymbol X)^2
P(z\mid\boldsymbol X)
\right\}
}_{V_{\mathrm{dist}}}\\
&\quad+
\underbrace{
E_{\boldsymbol X}\!\left\{
\sum_z\delta_{\mathrm{remain}}(z,\boldsymbol X)^2
P(z\mid\boldsymbol X)
\right\}
}_{V_{\mathrm{remain}}}\\
&\quad+
\underbrace{
2E_{\boldsymbol X}\!\left\{
\sum_z
\delta_{\mathrm{dist}}(z,\boldsymbol X)
\delta_{\mathrm{remain}}(z,\boldsymbol X)
P(z\mid\boldsymbol X)
\right\}
}_{V_{\mathrm{cross}}}\\
&=V_{\mathrm{dist}}+V_{\mathrm{remain}}+V_{\mathrm{cross}}.
\end{align*}

\section{Derivation of the Group Contrasts in the Dichotomous Case}
\label{sec:binary_group_contrasts}

Suppose \(A\in\{0,1\}\) and \(Z\in\{z,z^*\}\), where \(z\)
is the index group and \(z^*\) is the comparison group. For
\(a\in\{0,1\}\), write
\[
Y(a)=Y(0)+a\{Y(1)-Y(0)\}.
\]
By the law of total probability, the common reference
assignment distribution satisfies
\[
\begin{aligned}
P(A=a\mid\boldsymbol x)
&=P(Z=z\mid\boldsymbol x)P(A=a\mid z,\boldsymbol x)\\
&\quad+P(Z=z^*\mid\boldsymbol x)P(A=a\mid z^*,\boldsymbol x).
\end{aligned}
\]

\paragraph{Care-setting distribution contrast.}
Subtracting the common reference probability from the
index group's assignment probability gives
\[
\begin{aligned}
&P(A=a\mid z,\boldsymbol x)-P(A=a\mid\boldsymbol x)\\
&\quad=
P(Z=z^*\mid\boldsymbol x)
\bigl[P(A=a\mid z,\boldsymbol x)
      -P(A=a\mid z^*,\boldsymbol x)\bigr].
\end{aligned}
\]
Substituting this expression into the definition of
\(\delta_{\mathrm{dist}}\) yields
\[
\begin{aligned}
\delta_{\mathrm{dist}}(z,\boldsymbol x)
&=
P(Z=z^*\mid\boldsymbol x)
\sum_{a=0}^{1} E\{Y(a)\mid z,\boldsymbol x\}\\
&\qquad\times
\bigl[P(A=a\mid z,\boldsymbol x)
      -P(A=a\mid z^*,\boldsymbol x)\bigr].
\end{aligned}
\]
Writing \(Y(a)=Y(0)+a\{Y(1)-Y(0)\}\), the term involving
\(E\{Y(0)\mid z,\boldsymbol x\}\) cancels because the assignment
probability differences sum to zero. The term multiplied by
\(a\) contributes only at \(a=1\), giving
\[
\begin{aligned}
\delta_{\mathrm{dist}}(z,\boldsymbol x)
&=
P(Z=z^*\mid\boldsymbol x)
E\{Y(1)-Y(0)\mid z,\boldsymbol x\}\\
&\qquad\times
\bigl[P(A=1\mid z,\boldsymbol x)
      -P(A=1\mid z^*,\boldsymbol x)\bigr].
\end{aligned}
\]

\paragraph{Remaining standardized group contrast.}
Expanding the common reference distribution in the definition
of \(\delta_{\mathrm{remain}}\) gives
\[
\begin{aligned}
\delta_{\mathrm{remain}}(z,\boldsymbol x)
&=
\sum_{z'\in\{z,z^*\}}P(Z=z'\mid\boldsymbol x)
\sum_{a=0}^{1}P(A=a\mid z',\boldsymbol x)\\
&\qquad\times
\bigl[E\{Y(a)\mid z,\boldsymbol x\}
      -E\{Y(a)\mid z',\boldsymbol x\}\bigr].
\end{aligned}
\]
The term for \(z'=z\) is zero. For \(z'=z^*\), substituting
\(Y(a)=Y(0)+a\{Y(1)-Y(0)\}\) separates the baseline outcome
difference from the difference in average setting effects.
Summing over \(a\) leaves the baseline difference unchanged
because the assignment probabilities sum to one, and weights
the difference in setting effects by the comparison group's
probability of assignment to \(A=1\). Thus,
\[
\begin{aligned}
\delta_{\mathrm{remain}}(z,\boldsymbol x)
&=
P(Z=z^*\mid\boldsymbol x)\Bigl\{
E\{Y(0)\mid z,\boldsymbol x\}
-E\{Y(0)\mid z^*,\boldsymbol x\}\\
&\qquad+
P(A=1\mid z^*,\boldsymbol x)
\Bigl[
E\{Y(1)-Y(0)\mid z,\boldsymbol x\}\\
&\hspace{43mm}
-E\{Y(1)-Y(0)\mid z^*,\boldsymbol x\}
\Bigr]\Bigr\}.
\end{aligned}
\]

\section{Further Decomposition of $V_{\mathrm{remain}}$}
\label{sec:V_remain_further}

\subsection{Single Reference Setting}
Choose a baseline setting $a_0$. Using
$Y(a)=Y(a_0)+\{Y(a)-Y(a_0)\}$, we obtain
\begin{align*}
\delta_{\mathrm{remain}}(z,\boldsymbol x)
&=\sum_{z^*,a}
\left[
E\{Y(a)\mid z,\boldsymbol x\}
-E\{Y(a)\mid z^*,\boldsymbol x\}
\right]
P(a\mid z^*,\boldsymbol x)P(z^*\mid\boldsymbol x)\\
&=\delta_{\mathrm{baseline}}(z,\boldsymbol x)
+\delta_{\mathrm{effect}}(z,\boldsymbol x),
\end{align*}
where
\begin{align*}
\delta_{\mathrm{baseline}}(z,\boldsymbol x)
&=E\{Y(a_0)\mid z,\boldsymbol x\}
-\sum_{z^*}E\{Y(a_0)\mid z^*,\boldsymbol x\}
P(z^*\mid\boldsymbol x),\\
\delta_{\mathrm{effect}}(z,\boldsymbol x)
&=\sum_{z^*,a}
\Big[
E\{Y(a)-Y(a_0)\mid z,\boldsymbol x\}\\
&\hspace{25mm}
-E\{Y(a)-Y(a_0)\mid z^*,\boldsymbol x\}
\Big]
P(a\mid z^*,\boldsymbol x)P(z^*\mid\boldsymbol x).
\end{align*}
The baseline contrast compares group-specific expected outcomes
under $a_0$ with their population mean.
The effect contrast captures differences across groups in
average setting effects relative to $a_0$, weighted by
assignment probabilities and group prevalences.

Expanding the square gives
\begin{align*}
V_{\mathrm{remain}}
&=E_{\boldsymbol X}\!\left\{
\sum_z
\left[
\delta_{\mathrm{baseline}}(z,\boldsymbol X)
+\delta_{\mathrm{effect}}(z,\boldsymbol X)
\right]^2P(z\mid\boldsymbol X)
\right\}\\
&=V_{\mathrm{baseline}}
+V_{\mathrm{effect}}
+V_{\mathrm{baseline,effect}},
\end{align*}
where
\begin{align*}
V_{\mathrm{baseline}}
&=E_{\boldsymbol X}\!\left\{
\sum_z\delta_{\mathrm{baseline}}(z,\boldsymbol X)^2
P(z\mid\boldsymbol X)\right\},\\
V_{\mathrm{effect}}
&=E_{\boldsymbol X}\!\left\{
\sum_z\delta_{\mathrm{effect}}(z,\boldsymbol X)^2
P(z\mid\boldsymbol X)\right\},\\
V_{\mathrm{baseline,effect}}
&=2E_{\boldsymbol X}\!\left\{
\sum_z\delta_{\mathrm{baseline}}(z,\boldsymbol X)
\delta_{\mathrm{effect}}(z,\boldsymbol X)
P(z\mid\boldsymbol X)\right\}.
\end{align*}
In particular,
$$
V_{\mathrm{baseline}}
=E_{\boldsymbol X}\!\left[
V_{Z\mid\boldsymbol X}
\left\{E(Y(a_0)\mid Z,\boldsymbol X)\right\}
\right].
$$

When $A,Z\in\{0,1\}$, choose $a_0=0$ and let $z^*$ denote
the other group. The two contrasts simplify to
\begin{align*}
\delta_{\mathrm{baseline}}(z,\boldsymbol x)
&=P(Z=z^*\mid\boldsymbol x)
\left[
E\{Y(0)\mid z,\boldsymbol x\}
-E\{Y(0)\mid z^*,\boldsymbol x\}
\right],\\
\delta_{\mathrm{effect}}(z,\boldsymbol x)
&=P(Z=z^*\mid\boldsymbol x)
P(A=1\mid Z=z^*,\boldsymbol x)\\
&\quad\times
\left[
E\{Y(1)-Y(0)\mid z,\boldsymbol x\}
-E\{Y(1)-Y(0)\mid z^*,\boldsymbol x\}
\right].
\end{align*}
The separate components depend on the choice of baseline
setting, whereas their sum $V_{\mathrm{remain}}$ does not.
The cross-term can be positive or negative, reflecting
reinforcement or offsetting between the baseline and
effect-heterogeneity contrasts.

\subsection{Equal Weights Across Care Settings}

Rather than choosing a single baseline setting, define
\[
\bar Y=\frac{1}{J}\sum_{b=1}^{J}Y(b)
\]
as the equally weighted average of an individual's potential
outcomes across care settings. Its conditional mean serves
as an average-care reference, following the benchmarking
approach of \citet{varewyck2014shrinkage}, and also used in
the disparity decomposition by \citet{vanderlinde2026causal}.

Using
\[
Y(a)=\bar Y+\{Y(a)-\bar Y\},
\]
we obtain
\begin{align*}
\delta_{\mathrm{remain}}(z,\boldsymbol x)
&=
\sum_{z^*,a}
\left[
E\{Y(a)\mid z,\boldsymbol x\}
-
E\{Y(a)\mid z^*,\boldsymbol x\}
\right]
P(a\mid z^*,\boldsymbol x)P(z^*\mid\boldsymbol x)\\
&=
\delta_{\mathrm{baseline}}(z,\boldsymbol x)
+
\delta_{\mathrm{effect}}(z,\boldsymbol x),
\end{align*}
where
\begin{align*}
\delta_{\mathrm{baseline}}(z,\boldsymbol x)
&=
E(\bar Y\mid z,\boldsymbol x)
-
\sum_{z^*}
E(\bar Y\mid z^*,\boldsymbol x)
P(z^*\mid\boldsymbol x),\\
\delta_{\mathrm{effect}}(z,\boldsymbol x)
&=
\sum_{z^*,a}
\Big[
E\{Y(a)-\bar Y\mid z,\boldsymbol x\}\\
&\hspace{23mm}
-
E\{Y(a)-\bar Y\mid z^*,\boldsymbol x\}
\Big]
P(a\mid z^*,\boldsymbol x)P(z^*\mid\boldsymbol x).
\end{align*}
The baseline contrast compares group-specific expected outcomes
under the equally weighted average-care reference with their
population mean. The effect contrast summarizes group
differences in setting-specific deviations from this reference,
weighted by assignment probabilities and group prevalences.

Expanding the square gives
\[
V_{\mathrm{remain}}
=
V_{\mathrm{baseline}}
+
V_{\mathrm{effect}}
+
V_{\mathrm{baseline,effect}},
\]
where
\begin{align*}
V_{\mathrm{baseline}}
&=
E_{\boldsymbol X}\left[
\sum_z
\delta_{\mathrm{baseline}}(z,\boldsymbol X)^2
P(z\mid\boldsymbol X)
\right],\\
V_{\mathrm{effect}}
&=
E_{\boldsymbol X}\left[
\sum_z
\delta_{\mathrm{effect}}(z,\boldsymbol X)^2
P(z\mid\boldsymbol X)
\right],\\
V_{\mathrm{baseline,effect}}
&=
2E_{\boldsymbol X}\left[
\sum_z
\delta_{\mathrm{baseline}}(z,\boldsymbol X)
\delta_{\mathrm{effect}}(z,\boldsymbol X)
P(z\mid\boldsymbol X)
\right].
\end{align*}

\section{Variance and Covariance Representations of the Group Components
in the Absence of \texorpdfstring{\(A\times Z\)}{A by Z} Interaction}
\label{sec:V_cross}

In the absence of \(A\times Z\) interaction on the conditional potential-outcome
mean scale, the care-setting distribution contrast and the remaining
standardized group contrast each have mean zero across groups at every
allowable covariate profile. The three components of \(V_{\mathrm{group}}\)
then have a variance--covariance representation analogous to that derived by
\citet{chen2022causal} for between-hospital variation. In their setting,
hospital was the exposure. Here the contrasts are across sociodemographic groups.

At fixed \(\boldsymbol x\), \(\delta_{\mathrm{group}}(z,\boldsymbol x)\)
is the difference between the group-specific and population interventional
means and therefore averages to zero over \(Z\mid\boldsymbol x\).
Since \(\delta_{\mathrm{group}}=\delta_{\mathrm{dist}}+\delta_{\mathrm{remain}}\),
it suffices to show that \(\delta_{\mathrm{dist}}\) averages to zero as well.

Under the no-interaction condition, the mean effect of assignment to \(a\)
versus a baseline setting \(a_0\) is the same across groups at each
\(\boldsymbol x\). Thus,
\[
E\{Y(a)\mid z,\boldsymbol x\}
=
E\{Y(a_0)\mid z,\boldsymbol x\}
+
E\{Y(a)-Y(a_0)\mid\boldsymbol x\}.
\]
Substituting this expression into the conditional mean of
\(\delta_{\mathrm{dist}}\) and collecting terms gives
\begin{align*}
&E_{Z\mid\boldsymbol x}
\{\delta_{\mathrm{dist}}(Z,\boldsymbol x)\}\\
&\quad=
\sum_z E\{Y(a_0)\mid z,\boldsymbol x\}P(z\mid\boldsymbol x)
\left[
\sum_a P(a\mid z,\boldsymbol x)
-
\sum_a P(a\mid\boldsymbol x)
\right]\\
&\qquad+
\sum_a E\{Y(a)-Y(a_0)\mid\boldsymbol x\}
\left[
\sum_z P(a\mid z,\boldsymbol x)P(z\mid\boldsymbol x)
-
P(a\mid\boldsymbol x)
\right].
\end{align*}
The first bracket is zero because both assignment distributions
sum to one. The second is zero by the law of total probability.
Thus both contrasts have conditional mean zero. Evaluating the contrasts at
\((Z,\boldsymbol X)\), the three components defined in Appendix~\ref{sec:V_group_proof}
can therefore be written as
\begin{align*}
V_{\mathrm{dist}}
&= E_{\boldsymbol X}\!\left[
V_{Z\mid\boldsymbol X}\{\delta_{\mathrm{dist}}(Z,\boldsymbol X)\}\right],\\
V_{\mathrm{remain}}
&= E_{\boldsymbol X}\!\left[
V_{Z\mid\boldsymbol X}\{\delta_{\mathrm{remain}}(Z,\boldsymbol X)\}\right],\\
V_{\mathrm{cross}}
&= 2E_{\boldsymbol X}\!\left[
\operatorname{Cov}_{Z\mid\boldsymbol X}\{
\delta_{\mathrm{dist}}(Z,\boldsymbol X),\delta_{\mathrm{remain}}(Z,\boldsymbol X)\}\right].
\end{align*}
At each allowable covariate profile, the two variances measure how much the
respective contrasts vary across groups, and the covariance measures how they
vary together. Each is averaged over the distribution of \(\boldsymbol X\).

\section{Path-Specific and Interventional Interpretations of the Group Contrasts}
\label{sec:path_specific_effect}

\subsection{Path-Specific Effects with Allowable Covariates Held Fixed}
\label{sec:path_specific_fixed_x}

Consider the special case in which the set of non-allowable
covariates \(\boldsymbol N\) is empty and both \(U_1\) and \(U_2\)
are absent. Suppose that interventions on \(Z\) and
\(\boldsymbol X\) are well defined. We represent the DAG for
this special case, shown in Figure~\ref{fig:dag_intervene}, by
a nonparametric structural equation model with independent
errors (NPSEM-IE) \citep{pearl2009causality}.

\begin{figure*}[!htbp]
\centering
\begin{tikzpicture}[
    >=Latex,
    node distance=1 cm and 1 cm,
    semithick,
    state/.style={circle, draw, minimum width=0.7 cm}
]
    \node[state] (z) at (0,0) {$Z$};
    \node[state, right=of z] (a) {$A$};
    \node[state, right=of a] (y) {$Y$};
    \node[state, above=of a] (x) {$\boldsymbol X$};
    \node[state, below=of a] (tau) {$\tau$};
    \node[state, below=of y] (epsilon) {$\varepsilon$};
    \node[state, above=of z] (xi) {$\xi$};

    \path[->]
        (xi) edge (x)
        (tau) edge (a)
        (epsilon) edge (y)
        (z) edge (x)
        (z) edge (a)
        (z) edge[bend right=45] (y)
        (x) edge (a)
        (x) edge (y)
        (a) edge (y);
\end{tikzpicture}
\caption{Causal directed acyclic graph for the special case
with group membership \(Z\), allowable covariates
\(\boldsymbol X\), care-setting assignment \(A\), and outcome
\(Y\). The structural errors \(\xi\), \(\tau\), and
\(\varepsilon\) are mutually independent and independent of \(Z\).}
\label{fig:dag_intervene}
\end{figure*}
\FloatBarrier

Specifically, let \(\boldsymbol X=h(Z,\xi)\),
\(A=g(Z,\boldsymbol X,\tau)\), and
\(Y=f(Z,\boldsymbol X,A,\varepsilon)\), with
\(Z,\xi,\tau,\varepsilon\) mutually independent. Define
\[
A(z,\boldsymbol x)=g(z,\boldsymbol x,\tau)
\qquad\text{and}\qquad
Y(z,\boldsymbol x,a)=f(z,\boldsymbol x,a,\varepsilon).
\]
We evaluate the contrasts within the stratum
\(\boldsymbol X=\boldsymbol x\) and fix \(\boldsymbol X\) at
\(\boldsymbol x\) under both interventions on \(Z\).

Assume consistency and positivity for the group and care-setting
levels used in the comparisons. By consistency,
\(Y(a)=Y(z,\boldsymbol x,a)\) within the stratum
\((Z,\boldsymbol X)=(z,\boldsymbol x)\). The structural
independence assumptions then give
\begin{align*}
E\{Y(a)\mid z,\boldsymbol x\}
&=E\{Y(z,\boldsymbol x,a)\mid\boldsymbol x\},\\
P(a\mid z^*,\boldsymbol x)
&=P\{A(z^*,\boldsymbol x)=a\mid\boldsymbol x\}.
\end{align*}
They also imply the cross-world independence
\[
Y(z,\boldsymbol x,a)\independent A(z^*,\boldsymbol x)
\mid(\boldsymbol X=\boldsymbol x).
\]
Consequently, the expected outcome for group \(z\) under the
assignment distribution of group \(z^*\) can be written as
\begin{align*}
\mu(z,z^*,\boldsymbol x)
&=
\sum_a E\{Y(a)\mid z,\boldsymbol x\}
P(a\mid z^*,\boldsymbol x)\\
&=
\sum_a E\{Y(z,\boldsymbol x,a)\mid\boldsymbol x\}
P\{A(z^*,\boldsymbol x)=a\mid\boldsymbol x\}\\
&=
E\{Y(z,\boldsymbol x,A(z^*,\boldsymbol x))\mid\boldsymbol x\}.
\end{align*}

Using the definition of the common reference assignment
distribution in the main text, the care-setting distribution
contrast becomes
\begin{align*}
\delta_{\mathrm{dist}}(z,\boldsymbol x)
&=
\sum_{z^*}P(z^*\mid\boldsymbol x)
\left[
\mu(z,z,\boldsymbol x)-\mu(z,z^*,\boldsymbol x)
\right]\\
&=
\sum_{z^*}P(z^*\mid\boldsymbol x)
\Bigl[
E\{Y(z,\boldsymbol x,A(z,\boldsymbol x))\mid\boldsymbol x\}\\
&\hspace{33mm}-
E\{Y(z,\boldsymbol x,A(z^*,\boldsymbol x))\mid\boldsymbol x\}
\Bigr].
\end{align*}
For each comparison group \(z^*\), this changes the group input
to the assignment equation from \(z^*\) to \(z\) while holding
the group input to the outcome equation at \(z\).

Similarly, the remaining standardized group contrast becomes
\begin{align*}
\delta_{\mathrm{remain}}(z,\boldsymbol x)
&=
\sum_{z^*}P(z^*\mid\boldsymbol x)
\left[
\mu(z,z^*,\boldsymbol x)-\mu(z^*,z^*,\boldsymbol x)
\right]\\
&=
\sum_{z^*}P(z^*\mid\boldsymbol x)
\Bigl[
E\{Y(z,\boldsymbol x,A(z^*,\boldsymbol x))\mid\boldsymbol x\}\\
&\hspace{33mm}-
E\{Y(z^*,\boldsymbol x,A(z^*,\boldsymbol x))\mid\boldsymbol x\}
\Bigr].
\end{align*}
For each comparison group \(z^*\), this changes the group input
to the outcome equation from \(z^*\) to \(z\) while holding
assignment at \(A(z^*,\boldsymbol x)\).

With \(\boldsymbol X\) fixed at \(\boldsymbol x\),
\(\delta_{\mathrm{dist}}(z,\boldsymbol x)\) averages the pairwise effects
along \(Z\to A\to Y\), and \(\delta_{\mathrm{remain}}(z,\boldsymbol x)\)
averages the pairwise effects along \(Z\to Y\), in each case weighting
comparison groups by \(P(z^*\mid\boldsymbol x)\).
This interpretation follows the path-specific approach of
\citet{vanderweele2014effect}, with the allowable covariates set to
the same value under both group interventions.

\subsection{Care-Setting Interventions Within Observed Allowable Covariate Strata}

Suppose again that \(\boldsymbol N=\varnothing\). Consider patients in
group \(z\) with observed allowable covariate profile \(\boldsymbol x\).
For these patients, let \(G(z^*,\boldsymbol x)\) assign a care setting
by drawing from \(P(A=a\mid z^*,\boldsymbol x)\) independently of their
potential outcomes. Group membership and the allowable covariates retain
their observed values. Under consistency, positivity for the settings
used in the comparison, and \(Y(a)\independent A\mid(Z,\boldsymbol X)\),
the mean outcome under this assignment is
\begin{align*}
E\!\left[Y\{G(z^*,\boldsymbol x)\}\mid Z=z,\boldsymbol X=\boldsymbol x\right]
&=\sum_a E\{Y(a)\mid z,\boldsymbol x\}P(a\mid z^*,\boldsymbol x)\\
&=\sum_a E(Y\mid a,z,\boldsymbol x)P(a\mid z^*,\boldsymbol x)
=\mu(z,z^*,\boldsymbol x).
\end{align*}
Thus the same mean used in the preceding subsection has an interpretation
under an intervention on care-setting assignment alone.

For each comparison group \(z^*\), the difference
\(\mu(z,z,\boldsymbol x)-\mu(z,z^*,\boldsymbol x)\) compares two
assignment distributions for the same patients in group \(z\).
The difference \(\mu(z,z^*,\boldsymbol x)-\mu(z^*,z^*,\boldsymbol x)\)
compares patients from the two observed groups at the same profile when
both receive the assignment distribution of group \(z^*\).
Averaging these differences over \(z^*\) with weights
\(P(z^*\mid\boldsymbol x)\) gives \(\delta_{\mathrm{dist}}(z,\boldsymbol x)\)
and \(\delta_{\mathrm{remain}}(z,\boldsymbol x)\), respectively.
In this special case, applying a group's own assignment distribution
recovers its observed mean, \(\mu(z,z,\boldsymbol x)=E(Y\mid z,\boldsymbol x)\),
by the law of total expectation.

This gives an observed-group analogue of the randomized interventional
decomposition of \citet{vanderweele2014effect}. Their decomposition
defines effects of intervening on the exposure. Here only care-setting
assignment is randomized, and the two groups are compared at the same
observed \(\boldsymbol x\), which may itself be affected by \(Z\).
The remaining contrast therefore describes the disparity under a common
assignment policy. Interpreting it as a direct causal effect of \(Z\)
requires additional assumptions (as in~\ref{sec:path_specific_fixed_x}).

\subsection{Care-Setting Interventions in the Presence of Non-Allowable Covariates}

Now suppose that \(\boldsymbol N\) is present and may affect both
care-setting assignment and the outcome. As before, consider patients
in group \(z\) with observed allowable covariate profile \(\boldsymbol x\).
The draw \(G(z^*,\boldsymbol x)\) assigns a setting according to
\(P(A=a\mid z^*,\boldsymbol x)\), independently of these patients'
\(\boldsymbol N\) and potential outcomes. Under consistency, positivity,
and \(Y(a)\independent A\mid(Z,\boldsymbol X,\boldsymbol N)\),
the mean outcome under this assignment is
\begin{align*}
\mu(z,z^*,\boldsymbol x)
&=E\!\left[Y\{G(z^*,\boldsymbol x)\}\mid Z=z,\boldsymbol X=\boldsymbol x\right]\\
&=\sum_a P(a\mid z^*,\boldsymbol x)
  \int E(Y\mid a,z,\boldsymbol n,\boldsymbol x)
  \,dF(\boldsymbol n\mid z,\boldsymbol x).
\end{align*}
The assignment probability averages over the distribution of \(\boldsymbol N\)
in the comparison group \(z^*\) at \(\boldsymbol x\). The outcome integral
uses the distribution of \(\boldsymbol N\) in the recipient group \(z\)
at that same observed profile.

For each \(z^*\), \(\mu(z,z,\boldsymbol x)-\mu(z,z^*,\boldsymbol x)\)
compares assignment policies while retaining the recipient group's
distribution of \(\boldsymbol N\).
The difference \(\mu(z,z^*,\boldsymbol x)-\mu(z^*,z^*,\boldsymbol x)\)
compares the two groups under a common policy, allowing each group to retain
its own distribution of \(\boldsymbol N\). Averaging these differences with
weights \(P(z^*\mid\boldsymbol x)\) gives \(\delta_{\mathrm{dist}}\) and
\(\delta_{\mathrm{remain}}\), respectively. Even a draw from the recipient
group's own assignment distribution removes any dependence of assignment
on \(\boldsymbol N\) within that stratum. Consequently,
\(E(Y\mid z,\boldsymbol x)-\mu(z,z,\boldsymbol x)=D_N(z,\boldsymbol x)\)
can be nonzero (Appendix~\ref{app:observed-variance-N}).

When \(\boldsymbol X=\varnothing\), the policy draws from \(P(a\mid z^*)\)
and the outcome integral averages over \(F(\boldsymbol n\mid z)\).
This is the observed-group analogue of the randomized interventional
decomposition of \citet{vanderweele2014effect} with an exposure-induced
mediator--outcome confounder. An interpretation as interventional indirect
and direct effects of \(Z\) additionally requires well-defined interventions
on \(Z\) and exchangeability for its effects on assignment and outcome.

\section{Group Contrasts in the Presence of Unmeasured Characteristics}
\label{sec:unmeasured_u}

Conditioning on \(\boldsymbol X\) can induce associations between
group membership \(Z\) and unmeasured characteristics through
collider stratification. We examine how such an association can enter
the care-setting distribution contrast and the remaining standardized group
contrast within an allowable covariate profile. We consider the two scenarios in
Figure~\ref{fig:DAG_include_U} separately: assignment-related
selection through \(U_1\) in panel (b) and outcome-related
selection through \(U_2\) in panel (c), neither of which violates
conditional exchangeability \(Y(a)\perp A\mid(Z,\boldsymbol X,\boldsymbol N)\).

Let \(m_a(z,\boldsymbol x)=E\{Y(a)\mid z,\boldsymbol x\}\) denote the
mean potential outcome at setting \(a\), averaged over the natural
distribution of \(\boldsymbol N\) and the unmeasured characteristic
within group \(z\) at profile \(\boldsymbol x\). In both scenarios,
the identification assumptions in the main text give
\[
m_a(z,\boldsymbol x)
=\int E(Y\mid A=a,Z=z,\boldsymbol X=\boldsymbol x,\boldsymbol N=\boldsymbol n)
\,dF(\boldsymbol n\mid z,\boldsymbol x).
\]
The group contrasts can be written in pairwise form as
\begin{align*}
\delta_{\mathrm{dist}}(z,\boldsymbol x)
&=
\sum_{z^*,a}
m_a(z,\boldsymbol x)
\left[
P(a\mid z,\boldsymbol x)-P(a\mid z^*,\boldsymbol x)
\right]P(z^*\mid\boldsymbol x),\\
\delta_{\mathrm{remain}}(z,\boldsymbol x)
&=
\sum_{z^*,a}
\left[
m_a(z,\boldsymbol x)
-m_a(z^*,\boldsymbol x)
\right]
P(a\mid z^*,\boldsymbol x)P(z^*\mid\boldsymbol x).
\end{align*}

\paragraph{Assignment-related selection through \(U_1\).}
In panel (b), we examine the contribution through assignment
probabilities, taking the means \(m_a(z,\boldsymbol x)\) as given.
Assignment probabilities can depend on \(U_1\), so differences in
\(F(u_1\mid z,\boldsymbol x)\) across groups can contribute to differences in
\[
P(a\mid z,\boldsymbol x)
=\int P(a\mid z,\boldsymbol x,u_1)\,dF(u_1\mid z,\boldsymbol x).
\]
These differences can enter \(\delta_{\mathrm{dist}}\) through the
assignment contrast in the first expression above. For each comparison
group \(z^*\), the second expression averages the group difference
\(m_a(z,\boldsymbol x)-m_a(z^*,\boldsymbol x)\) over settings with weights
\(P(a\mid z^*,\boldsymbol x)\). In the absence of an additive \(A\times Z\)
interaction in these averaged means, the group difference is constant
across settings. The weights then sum to one and the contribution through
assignment probabilities is confined to \(\delta_{\mathrm{dist}}\).
When the group difference varies across settings, \(U_1\) can contribute
through the assignment probabilities to both contrasts.

\paragraph{Outcome-related selection through \(U_2\).}
In panel (c), we examine the contribution of \(U_2\) through the
potential-outcome means, taking the assignment probabilities as given.
For illustration, suppose that the averaged mean has the form
\[
m_a(z,\boldsymbol x)
=
\beta_0+\boldsymbol\beta_X^{\mathsf T}\boldsymbol x
+\alpha_a+\gamma_z+\eta_{az}
+(\beta_U+\lambda_a)E(U_2\mid z,\boldsymbol x).
\]
The last term arises by averaging an additive contribution
\((\beta_U+\lambda_a)U_2\) within \((z,\boldsymbol x)\).
Here \(\alpha_a\) and \(\gamma_z\) are the setting and group main effects,
and \(\eta_{az}\) is their interaction term. With \(\lambda_{a_0}=0\)
at the reference setting \(a_0\), \(\beta_U\) is the coefficient of \(U_2\)
at that setting and \(\lambda_a\) is the change in that coefficient at setting \(a\).
Substituting this expression into the two contrasts yields
\begin{align*}
\delta_{\mathrm{dist}}(z,\boldsymbol x)
&=
\sum_{z^*,a}
\left[
\alpha_a+\eta_{az}+\lambda_a E(U_2\mid z,\boldsymbol x)
\right] \left[
P(a\mid z,\boldsymbol x)-P(a\mid z^*,\boldsymbol x)
\right]P(z^*\mid\boldsymbol x),\\[4pt]
\delta_{\mathrm{remain}}(z,\boldsymbol x)
&=
\gamma_z-\sum_{z^*}\gamma_{z^*}P(z^*\mid\boldsymbol x)\\
&\quad+
\sum_{z^*,a}
(\eta_{az}-\eta_{az^*})
P(a\mid z^*,\boldsymbol x)P(z^*\mid\boldsymbol x)\\
&\quad+
\beta_U\left[
E(U_2\mid z,\boldsymbol x)-E(U_2\mid\boldsymbol x)
\right]\\
&\quad+
\sum_{z^*,a}\lambda_a
\left[
E(U_2\mid z,\boldsymbol x)-E(U_2\mid z^*,\boldsymbol x)
\right] P(a\mid z^*,\boldsymbol x)P(z^*\mid\boldsymbol x).
\end{align*}
The displayed expression for \(\delta_{\mathrm{dist}}\) omits the
\(\beta_U E(U_2\mid z,\boldsymbol x)\) term from \(m_a(z,\boldsymbol x)\)
because, for each comparison group \(z^*\), its contribution is
\[
\beta_U E(U_2\mid z,\boldsymbol x)
\sum_a \big[P(a\mid z,\boldsymbol x)-P(a\mid z^*,\boldsymbol x)\big]
=0.
\]
Both assignment distributions sum to one. If the \(A\times U_2\) interaction is absent, the
contribution of $U_2$ through the outcome mean is therefore confined to
\(
\beta_U\left[
E(U_2\mid z,\boldsymbol x)-E(U_2\mid\boldsymbol x)
\right]
\)
in \(\delta_{\mathrm{remain}}\). If the \(A\times U_2\)
interaction is present, the terms involving \(\lambda_a\) can
contribute to both contrasts. In panel (c), these conclusions
about the outcome-mean contribution hold with or without an
\(A\times Z\) interaction.
Table~\ref{tab:unmeasured_selection} summarizes how \(U_1\) and \(U_2\)
can contribute to the two group contrasts under the interaction patterns considered above.

\begin{table}[!htbp]
\centering
\small
\caption{Possible contributions of unmeasured characteristics
to the group contrasts through assignment probabilities or potential-outcome means.}
\label{tab:unmeasured_selection}
\begin{tabularx}{\linewidth}{
@{}>{\raggedright\arraybackslash}Xcccc@{}
}
\toprule
Source of contribution
& \(A\times Z\)
& \(A\times U\)
& \(\delta_{\mathrm{dist}}\)
& \(\delta_{\mathrm{remain}}\) \\
\midrule
Panel (b): \(U_1\) through assignment probabilities
& Absent & --- & Possible & Zero \\
Panel (b): \(U_1\) through assignment probabilities
& Present & --- & Possible & Possible \\
Panel (c): \(U_2\) through potential-outcome means
& Either & Absent & Zero & Possible \\
Panel (c): \(U_2\) through potential-outcome means
& Either & Present & Possible & Possible \\
\bottomrule
\end{tabularx}
\par\smallskip
\begin{minipage}{\linewidth}
\footnotesize
In panel (b), \(A\times Z\) refers to interaction in the averaged means
\(m_a(z,\boldsymbol x)\). A dash indicates that no restriction on
\(A\times U_1\) is required for this assignment-weight argument.
In panel (c), the interaction terms are \(\eta_{az}\) and \(\lambda_a\)
in the illustrative model, and the assignment probabilities are held fixed.
``Zero'' refers to the
contribution through the indicated route.
\end{minipage}
\end{table}
\FloatBarrier

\section{Decomposition of $V_{\mathrm{setting}}$}
\label{sec:V_setting_proof}

Under the stochastic intervention,
$E(Y^{\mathcal G}\mid z,\boldsymbol x,G=a)
=E\{Y(a)\mid z,\boldsymbol x\}$.
The categorical variance identity therefore gives
\begin{align*}
&V_{G\mid z,\boldsymbol x}
\left\{E(Y^{\mathcal G}\mid z,\boldsymbol x,G)\right\}\\
&\quad=
\sum_a E\{Y(a)\mid z,\boldsymbol x\}^2
P(a\mid z,\boldsymbol x)
-\left[
\sum_a E\{Y(a)\mid z,\boldsymbol x\}
P(a\mid z,\boldsymbol x)
\right]^2\\
&\quad=
\frac12\sum_{a,a^*}
P(a\mid z,\boldsymbol x)P(a^*\mid z,\boldsymbol x)
\left[
E\{Y(a)\mid z,\boldsymbol x\}
-E\{Y(a^*)\mid z,\boldsymbol x\}
\right]^2\\
&\quad=
\sum_{a<a^*}
\pi_{aa^*}(z,\boldsymbol x)
\delta_{aa^*}(z,\boldsymbol x)^2.
\end{align*}
Hence,
$$
V_{\mathrm{setting}}
=
E_{\boldsymbol X}\!\left\{
\sum_{a<a^*}
E_{Z\mid\boldsymbol X}
\left[
\pi_{aa^*}(Z,\boldsymbol X)
\delta_{aa^*}(Z,\boldsymbol X)^2
\right]
\right\}.
$$
Here, for each pair and covariate profile, the covariance identity
and second-moment decomposition yield
\begin{align*}
&E_{Z\mid\boldsymbol X}
\left[
\pi_{aa^*}(Z,\boldsymbol X)
\delta_{aa^*}(Z,\boldsymbol X)^2
\right]\\
&\quad=
E_{Z\mid\boldsymbol X}
\{\pi_{aa^*}(Z,\boldsymbol X)\}
E_{Z\mid\boldsymbol X}
\{\delta_{aa^*}(Z,\boldsymbol X)^2\}\\
&\qquad+
\operatorname{Cov}_{Z\mid\boldsymbol X}
\left(
\pi_{aa^*}(Z,\boldsymbol X),
\delta_{aa^*}(Z,\boldsymbol X)^2
\right)\\
&\quad=
E_{Z\mid\boldsymbol X}
\{\pi_{aa^*}(Z,\boldsymbol X)\}
\left[
E_{Z\mid\boldsymbol X}
\{\delta_{aa^*}(Z,\boldsymbol X)\}
\right]^2\\
&\qquad+
E_{Z\mid\boldsymbol X}
\{\pi_{aa^*}(Z,\boldsymbol X)\}
V_{Z\mid\boldsymbol X}
\{\delta_{aa^*}(Z,\boldsymbol X)\}\\
&\qquad+
\operatorname{Cov}_{Z\mid\boldsymbol X}
\left(
\pi_{aa^*}(Z,\boldsymbol X),
\delta_{aa^*}(Z,\boldsymbol X)^2
\right).
\end{align*}
Finally, summing over unordered setting pairs and averaging over
$\boldsymbol X$ gives
\begin{align*}
V_{\mathrm{setting}}
&=V_{\mathrm{me}}+V_{\mathrm{em}}+V_{\mathrm{cov}},
\end{align*}
where
\begin{align*}
V_{\mathrm{me}}
&=E_{\boldsymbol X}\!\left\{
\sum_{a<a^*}
E_{Z\mid\boldsymbol X}
\{\pi_{aa^*}(Z,\boldsymbol X)\}
\left[
E_{Z\mid\boldsymbol X}
\{\delta_{aa^*}(Z,\boldsymbol X)\}
\right]^2
\right\},\\
V_{\mathrm{em}}
&=E_{\boldsymbol X}\!\left\{
\sum_{a<a^*}
E_{Z\mid\boldsymbol X}
\{\pi_{aa^*}(Z,\boldsymbol X)\}
V_{Z\mid\boldsymbol X}
\{\delta_{aa^*}(Z,\boldsymbol X)\}
\right\},\\
V_{\mathrm{cov}}
&=E_{\boldsymbol X}\!\left\{
\sum_{a<a^*}
\operatorname{Cov}_{Z\mid\boldsymbol X}
\left(
\pi_{aa^*}(Z,\boldsymbol X),
\delta_{aa^*}(Z,\boldsymbol X)^2
\right)
\right\}.
\end{align*}

\section{Model-based Estimators of Variance Components}
\label{sec:model_based_estimator}

Let
$\{(Y_i,A_i,Z_i,\boldsymbol N_i,\boldsymbol X_i)\}_{i=1}^n$
denote the observed sample.
We construct plug-in estimators using the fitted outcome,
assignment, group-membership, and non-allowable covariate
models, and average over the empirical distribution of
$\boldsymbol X$.

\subsection{Fitted Conditional Moments and Probabilities}

The fitted joint distribution of the non-allowable covariates is
$$
\widehat f_N(\boldsymbol n\mid z,\boldsymbol x)
=
\prod_{\ell=1}^{L}
f\!\left(
n_\ell\mid n_1,\ldots,n_{\ell-1},z,\boldsymbol x;
\widehat{\boldsymbol\beta}_\ell
\right).
$$
Define the estimated conditional potential-outcome means,
marginal assignment probabilities, and group-membership probabilities as
\begin{align*}
\widehat m_a(z,\boldsymbol x)
&=
\int
E(Y\mid a,z,\boldsymbol n,\boldsymbol x;
\widehat{\boldsymbol\theta})
\widehat f_N(\boldsymbol n\mid z,\boldsymbol x)
\,d\boldsymbol n,\\
\widehat g_a(z,\boldsymbol x)
&=
\int
P(A=a\mid z,\boldsymbol n,\boldsymbol x;
\widehat{\boldsymbol\eta})
\widehat f_N(\boldsymbol n\mid z,\boldsymbol x)
\,d\boldsymbol n,\\
\widehat p_z(\boldsymbol x)
&=
P(Z=z\mid\boldsymbol x;\widehat{\boldsymbol\phi}).
\end{align*}
Integrals are replaced by sums for discrete covariates and
evaluated numerically when necessary for continuous covariates.
The outcome means and assignment probabilities are
marginalized separately over the same fitted distribution
$\widehat f_N(\boldsymbol n\mid z,\boldsymbol x)$.
For mixed-effects outcome models, predictions retain the
fitted setting-specific random effects.

The estimated group-specific and population means under
$\mathcal G$ are
\begin{align*}
\widehat\mu_{\mathcal G}(z,\boldsymbol x)
&=
\sum_a
\widehat m_a(z,\boldsymbol x)
\widehat g_a(z,\boldsymbol x),\\
\widehat\mu_{\mathcal G}(\boldsymbol x)
&=
\sum_z
\widehat\mu_{\mathcal G}(z,\boldsymbol x)
\widehat p_z(\boldsymbol x).
\end{align*}
The estimated common reference assignment distribution is
$$
\widehat{\overline p}_a(\boldsymbol x)
=
\sum_z
\widehat g_a(z,\boldsymbol x)
\widehat p_z(\boldsymbol x).
$$

\subsection{Group-level Component Estimators}
\label{sec:group_components_estimators}

For each $(z,\boldsymbol x)$, define
\begin{align*}
\widehat\delta_{\mathrm{dist}}(z,\boldsymbol x)
&=
\sum_a
\widehat m_a(z,\boldsymbol x)
\left\{
\widehat g_a(z,\boldsymbol x)
-\widehat{\overline p}_a(\boldsymbol x)
\right\},\\
\widehat\delta_{\mathrm{remain}}(z,\boldsymbol x)
&=
\sum_a
\widehat m_a(z,\boldsymbol x)
\widehat{\overline p}_a(\boldsymbol x)
-\widehat\mu_{\mathcal G}(\boldsymbol x).
\end{align*}
The population benchmark in the remaining contrast is
$\widehat\mu_{\mathcal G}(\boldsymbol x)$, not the population
mean obtained after applying the common reference intervention
to all groups. The three group-level component estimators are
\begin{align*}
\widehat V_{\mathrm{dist}}
&=
\frac1n\sum_{i=1}^n\sum_z
\widehat\delta_{\mathrm{dist}}(z,\boldsymbol X_i)^2
\widehat p_z(\boldsymbol X_i),\\
\widehat V_{\mathrm{remain}}
&=
\frac1n\sum_{i=1}^n\sum_z
\widehat\delta_{\mathrm{remain}}(z,\boldsymbol X_i)^2
\widehat p_z(\boldsymbol X_i),\\
\widehat V_{\mathrm{cross}}
&=
\frac2n\sum_{i=1}^n\sum_z
\widehat\delta_{\mathrm{dist}}(z,\boldsymbol X_i)
\widehat\delta_{\mathrm{remain}}(z,\boldsymbol X_i)
\widehat p_z(\boldsymbol X_i).
\end{align*}

\subsection{Setting-level Component Estimators}
\label{sec:setting_components_estimators}

For each unordered pair $(a,a^*)$, define
\begin{align*}
\widehat\delta_{aa^*}(z,\boldsymbol x)
&=
\widehat m_a(z,\boldsymbol x)
-\widehat m_{a^*}(z,\boldsymbol x),\\
\widehat\pi_{aa^*}(z,\boldsymbol x)
&=
\widehat g_a(z,\boldsymbol x)
\widehat g_{a^*}(z,\boldsymbol x).
\end{align*}
Then
\begin{align*}
\widehat V_{\mathrm{me}}
&=
\frac1n\sum_{i=1}^n\sum_{a<a^*}
\left\{
\sum_z
\widehat\pi_{aa^*}(z,\boldsymbol X_i)
\widehat p_z(\boldsymbol X_i)
\right\}
\left\{
\sum_z
\widehat\delta_{aa^*}(z,\boldsymbol X_i)
\widehat p_z(\boldsymbol X_i)
\right\}^2,\\
\widehat V_{\mathrm{em}}
&=
\frac1n\sum_{i=1}^n\sum_{a<a^*}
\left\{
\sum_z
\widehat\pi_{aa^*}(z,\boldsymbol X_i)
\widehat p_z(\boldsymbol X_i)
\right\}\\
&\qquad\times
\sum_z
\left[
\widehat\delta_{aa^*}(z,\boldsymbol X_i)
-\sum_{z^*}
\widehat\delta_{aa^*}(z^*,\boldsymbol X_i)
\widehat p_{z^*}(\boldsymbol X_i)
\right]^2
\widehat p_z(\boldsymbol X_i),\\
\widehat V_{\mathrm{cov}}
&=
\frac1n\sum_{i=1}^n\sum_{a<a^*}
\Bigg[
\sum_z
\widehat\pi_{aa^*}(z,\boldsymbol X_i)
\widehat\delta_{aa^*}(z,\boldsymbol X_i)^2
\widehat p_z(\boldsymbol X_i)\\
&\qquad-
\left\{
\sum_z
\widehat\pi_{aa^*}(z,\boldsymbol X_i)
\widehat p_z(\boldsymbol X_i)
\right\}
\left\{
\sum_z
\widehat\delta_{aa^*}(z,\boldsymbol X_i)^2
\widehat p_z(\boldsymbol X_i)
\right\}
\Bigg].
\end{align*}

\subsection{Allowable-covariate and Residual Component Estimators}
\label{sec:case_mix_estimators}

Let
\[
\overline{\widehat\mu}_{\mathcal G}
=
\frac{1}{n}\sum_{i=1}^n
\widehat\mu_{\mathcal G}(\boldsymbol X_i).
\]
The allowable-covariate component is estimated by
\[
\widehat V_{\boldsymbol X}
=
\frac{1}{n}\sum_{i=1}^n
\left\{
\widehat\mu_{\mathcal G}(\boldsymbol X_i)
-\overline{\widehat\mu}_{\mathcal G}
\right\}^2.
\]

For each setting \(a\), let \(\widehat v_a(z,\boldsymbol x)\) denote the
estimated conditional variance of \(Y(a)\) given \(Z=z\) and
\(\boldsymbol X=\boldsymbol x\).
The residual component is then estimated by
\[
\widehat V_{\mathrm{residual}}
=
\frac{1}{n}\sum_{i=1}^n\sum_{z,a}
\widehat v_a(z,\boldsymbol X_i)
\widehat g_a(z,\boldsymbol X_i)
\widehat p_z(\boldsymbol X_i).
\]
For a binary outcome, we take
\[
\widehat v_a(z,\boldsymbol x)
=
\widehat m_a(z,\boldsymbol x)
\left\{1-\widehat m_a(z,\boldsymbol x)\right\}.
\]
For a continuous outcome, let \(\sigma_Y^2\) denote the constant
conditional variance given \((A,Z,\boldsymbol N,\boldsymbol X)\).
For the Gaussian fixed-effects model, we use the maximum
likelihood estimate
\[
\widehat\sigma_Y^2
=
\frac{1}{n}\sum_{i=1}^n
\left\{
Y_i-E(Y\mid A_i,Z_i,\boldsymbol N_i,\boldsymbol X_i;
\widehat{\boldsymbol\theta})
\right\}^2.
\]
For the Gaussian mixed-effects model, we use the residual
variance estimate from the maximum likelihood fit, with outcome
predictions conditional on the fitted setting-specific random
effects. For random forest and XGBoost in the continuous-outcome
simulations, we use the pooled out-of-fold mean squared prediction
error for the selected hyperparameters as a working residual
variance. The law of total variance then gives
\begin{align*}
\widehat v_a(z,\boldsymbol x)
&=
\widehat\sigma_Y^2
+\int
\left\{
E(Y\mid a,z,\boldsymbol n,\boldsymbol x;
\widehat{\boldsymbol\theta})
\right\}^2
\widehat f_N(\boldsymbol n\mid z,\boldsymbol x)
\,d\boldsymbol n\\
&\quad-\widehat m_a(z,\boldsymbol x)^2.
\end{align*}
The integral minus \(\widehat m_a(z,\boldsymbol x)^2\) is the
variance of the fitted outcome mean over the conditional
distribution \(\widehat F_N(\cdot\mid z,\boldsymbol x)\),
with the outcome prediction evaluated at setting a. Under heteroscedasticity,
we can replace \(\widehat\sigma_Y^2\) with the fitted conditional variance
averaged over that same distribution of \(\boldsymbol N\).

\subsection{Total Interventional Variance and Proportional Components}

The total interventional variance is estimated by
$$
\widehat V(Y^{\mathcal G})
=
\frac1n\sum_{i=1}^n\sum_{z,a}
\left\{\widehat v_a(z,\boldsymbol X_i)+\widehat m_a(z,\boldsymbol X_i)^2\right\}
\widehat g_a(z,\boldsymbol X_i)
\widehat p_z(\boldsymbol X_i)
-
\left(\overline{\widehat\mu}_{\mathcal G}\right)^2.
$$
These plug-in estimators satisfy
\begin{align*}
\widehat V(Y^{\mathcal G})
={}&
\widehat V_{\boldsymbol X}
+\widehat V_{\mathrm{dist}}
+\widehat V_{\mathrm{remain}}
+\widehat V_{\mathrm{cross}}\\
&+
\widehat V_{\mathrm{me}}
+\widehat V_{\mathrm{em}}
+\widehat V_{\mathrm{cov}}
+\widehat V_{\mathrm{residual}},
\end{align*}
up to numerical error. For each component $k$, its estimated percentage contribution is
$$
100\times
\frac{\widehat V_k}{\widehat V(Y^{\mathcal G})},
\qquad
\widehat V(Y^{\mathcal G})>0.
$$
The same calculations are repeated for each Bayesian bootstrap or parameter draw,
using its corresponding fitted quantities and total interventional variance.

\section{Data-Generating Mechanism for the Simulation Study}
\label{sec:simulation_dgm}

The \(J=5\) and \(J=10\) scenarios shared the same group and
covariate distributions and the same coefficients for settings
\(1,\ldots,5\). Setting \(A=1\) and group \(Z=1\) served as
reference categories.

\subsection{Group Membership and Covariates}
\label{sec:AF_simulation_race_coef}

We generated \(Z\in\{1,2,3\}\) with probabilities
\[
\bigl(P(Z=1),P(Z=2),P(Z=3)\bigr)=(0.45,0.35,0.20).
\]
The continuous non-allowable covariate \(N\) was generated from
\[
N\mid Z=z
\sim
\mathcal N\!\left(
-0.25+0.35\mathbb I(z=2)+0.70\mathbb I(z=3),\,0.70^2
\right).
\]
Conditional on \((Z,N)\), the allowable covariates \(X_1\) and
\(X_2\) were generated independently according to
\begin{align}
\operatorname{logit}P(X_1=1\mid Z=z,N=n)
&=
-0.10+0.10\mathbb I(z=2)+0.20\mathbb I(z=3)+0.30n,
\notag\\
X_2\mid Z=z,N=n
&\sim
\mathcal N\!\left(
0.60+0.10\mathbb I(z=2)+0.20\mathbb I(z=3)+0.35n,\,0.55^2
\right).
\label{eq:supp-simulation-covariates}
\end{align}

\subsection{Care-Setting Assignment}
\label{sec:AG_simulation_hosp_coef}

Let \(\boldsymbol x=(x_1,x_2)^{\mathsf T}\).
Care-setting assignment was generated with probabilities
\[
P(A=a\mid Z=z,N=n,\boldsymbol X=\boldsymbol x)
=
\frac{\exp\{c_a(z,n,\boldsymbol x)\}}
     {\sum_{j=1}^{J}\exp\{c_j(z,n,\boldsymbol x)\}},
\qquad a=1,\ldots,J.
\]
We set \(c_1(z,n,\boldsymbol x)=0\). For \(a=2,\ldots,J\),
the linear predictor was
\[
c_a(z,n,\boldsymbol x)
=
\eta_{0a}+\eta_{Xa,1}x_1+\eta_{Xa,2}x_2+\eta_{Na}n
+\eta_{a2}\mathbb I(z=2)+\eta_{a3}\mathbb I(z=3).
\]
Here \(\boldsymbol\eta_{Xa}=(\eta_{Xa,1},\eta_{Xa,2})^{\mathsf T}\).
The coefficients are given in
Table~\ref{tab:simulation_assignment_coef}.

\begin{table}[!htbp]
\centering
\caption{Coefficients for the care-setting assignment model.
The \(J=5\) scenario uses settings \(a=2,\ldots,5\) from the
table, and the \(J=10\) scenario uses all listed settings.}
\label{tab:simulation_assignment_coef}
\begin{tabular}{rrrrrrr}
\toprule
\(a\) & \(\eta_{0a}\) & \(\eta_{Xa,1}\) & \(\eta_{Xa,2}\)
& \(\eta_{Na}\) & \(\eta_{a2}\) & \(\eta_{a3}\) \\
\midrule
2  &  0.1 & 0.1 & 0.2 &  0.15 &  1.0 & 0.1 \\
3  & -0.2 & 0.5 & 0.5 & -0.10 &  0.2 & 0.5 \\
4  &  0.1 & 0.1 & 0.2 &  0.12 &  0.3 & 1.2 \\
5  &  0.1 & 0.5 & 0.5 & -0.08 & -0.4 & 0.7 \\
6  &  0.1 & 0.1 & 0.2 &  0.15 &  1.0 & 0.1 \\
7  & -0.2 & 0.5 & 0.5 & -0.10 &  0.2 & 0.5 \\
8  &  0.1 & 0.1 & 0.2 &  0.12 &  0.3 & 1.2 \\
9  &  0.1 & 0.5 & 0.5 & -0.08 & -0.4 & 0.7 \\
10 &  0.2 & 0.3 & 0.3 &  0.10 &  0.6 & 0.6 \\
\bottomrule
\end{tabular}
\end{table}

\subsection{Outcome Generation}
\label{sec:AI_simulation_outcome_coef}

The binary and continuous outcomes shared the linear predictor
\begin{align*}
L(a,z,n,\boldsymbol x)
&=
-0.5+1.4x_1-1.4x_2+0.30n
-0.5\mathbb I(z=2)+\mathbb I(z=3)\\
&\quad+
\theta_4^a+\theta_5^{a,2}\mathbb I(z=2)
+\theta_5^{a,3}\mathbb I(z=3)+\kappa_a n.
\end{align*}
Table~\ref{tab:simulation_outcome_coef} gives the care-setting
main-effect coefficients \(\theta_4^a\), the \(A\times Z\)
interaction coefficients \(\theta_5^{a,2}\) and \(\theta_5^{a,3}\), and
the \(A\times N\) interaction coefficients \(\kappa_a\).

Given \((A,Z,N,\boldsymbol X)=(a,z,n,\boldsymbol x)\), the
binary outcome followed a Bernoulli distribution with success
probability \(\operatorname{expit}\{L(a,z,n,\boldsymbol x)\}\).
The continuous outcome followed a Gaussian distribution with
mean \(L(a,z,n,\boldsymbol x)\) and variance \(1\).

\begin{table}[!htbp]
\centering
\caption{Care-setting main-effect and interaction coefficients
for both outcome types. The \(J=5\) scenario uses settings
\(a=1,\ldots,5\), and the \(J=10\) scenario uses all listed
settings.}
\label{tab:simulation_outcome_coef}
\begin{tabular}{rrrrr}
\toprule
\(a\) & \(\theta_4^a\) & \(\theta_5^{a,2}\) & \(\theta_5^{a,3}\) & \(\kappa_a\) \\
\midrule
1  & 0.0 &  0.0 &  0.0 &  0.00 \\
2  & 1.5 &  1.0 & -1.4 &  0.15 \\
3  & 0.5 &  1.2 &  1.5 & -0.10 \\
4  & 1.6 & -0.2 & -1.6 &  0.12 \\
5  & 0.0 & -0.5 & -0.3 & -0.06 \\
6  & 1.3 &  1.2 & -1.4 &  0.15 \\
7  & 0.0 &  1.3 &  1.3 & -0.10 \\
8  & 1.4 & -0.5 & -1.5 &  0.12 \\
9  & 0.0 & -0.2 &  1.6 & -0.06 \\
10 & 1.2 & -0.2 & -0.4 &  0.08 \\
\bottomrule
\end{tabular}
\end{table}

\section{Random forest and XGBoost hyperparameter tuning}
\label{sec:AH_hyperparam_tune}

\paragraph{Simulation study.}
Within each simulation replicate, we tuned random forest and XGBoost
by grid search using the same five cross-validation folds for both
learners. We selected hyperparameters by maximizing the mean validation
AUC for a dichotomous outcome or minimizing the mean validation RMSE
for a continuous outcome. We then refitted each learner on the full
replicate using the selected hyperparameters.

Random forest used 500 trees, with Gini splitting for dichotomous
outcomes and variance splitting for continuous outcomes. We varied
\texttt{mtry} from 1 to \(J+4\), the number of predictor columns after
encoding the categorical variables. This gave ranges of 1 to 9 for
\(J=5\) and 1 to 14 for \(J=10\). The candidate proportions \(p\) for
the minimum node size were
\(\{0,0.001,0.02,0.03,0.4,0.5,0.7\}\). For each fit, we set
\texttt{min.node.size} to
\(\max\{1,\lceil p n_{\mathrm{train}}\rceil\}\), where
\(n_{\mathrm{train}}\) is the number of observations used to fit
the model. XGBoost used logistic loss for dichotomous outcomes
and squared-error loss for continuous outcomes.

\paragraph{SEER application.}
Within each Bayesian bootstrap draw, we tuned both learners using
the same five outcome-stratified cross-validation folds. We used
the draw's patient weights for model fitting and validation,
selected hyperparameters by maximizing the mean weighted validation
AUC, and refitted each learner using all patients and their weights
for that draw.

Random forest used 500 trees and Gini splitting. We varied
\texttt{mtry} from 1 to 21, the number of encoded predictors.
The candidate node-size proportions were
\(\{0,0.02,0.03,0.4,0.5,0.7\}\), with
\texttt{min.node.size} set to
\(\max\{1,\lfloor p n_{\mathrm{train}}\rfloor\}\).
XGBoost used logistic loss. Table~\ref{tab:xgboost_search_grids} gives
the XGBoost search grids for the two analyses.
\begin{table}[!htbp]
\centering
\caption{XGBoost hyperparameter search grids for the simulation study and SEER application.}
\label{tab:xgboost_search_grids}
\small
\begin{tabular}{lcc}
\hline
Hyperparameter & Simulation study & SEER application \\
\hline
\texttt{nrounds}             & \(50,100,150\)         & \(50,100,150\) \\
\texttt{max\_depth}          & \(1,2,3\)              & \(1,2,3\) \\
\texttt{eta}                 & \(0.2,0.3,0.4,0.5,0.6,0.7\) & \(0.3,0.4\) \\
\texttt{gamma}               & \(0,0.6\)              & \(0\) \\
\texttt{colsample\_bytree}    & \(0.6,0.8\)            & \(0.6,0.8\) \\
\texttt{min\_child\_weight}   & \(1\)                  & \(1\) \\
\texttt{subsample}           & \(0.5,0.7,0.75,1\)    & \(0.5,0.75,1\) \\
\hline
\end{tabular}
\end{table}
\FloatBarrier

\section{Simulation results for a continuous outcome}\label{sec:AJ_cont_res}
\begin{figure}[!htbp]
    \centering
    \includegraphics[width=1\linewidth]{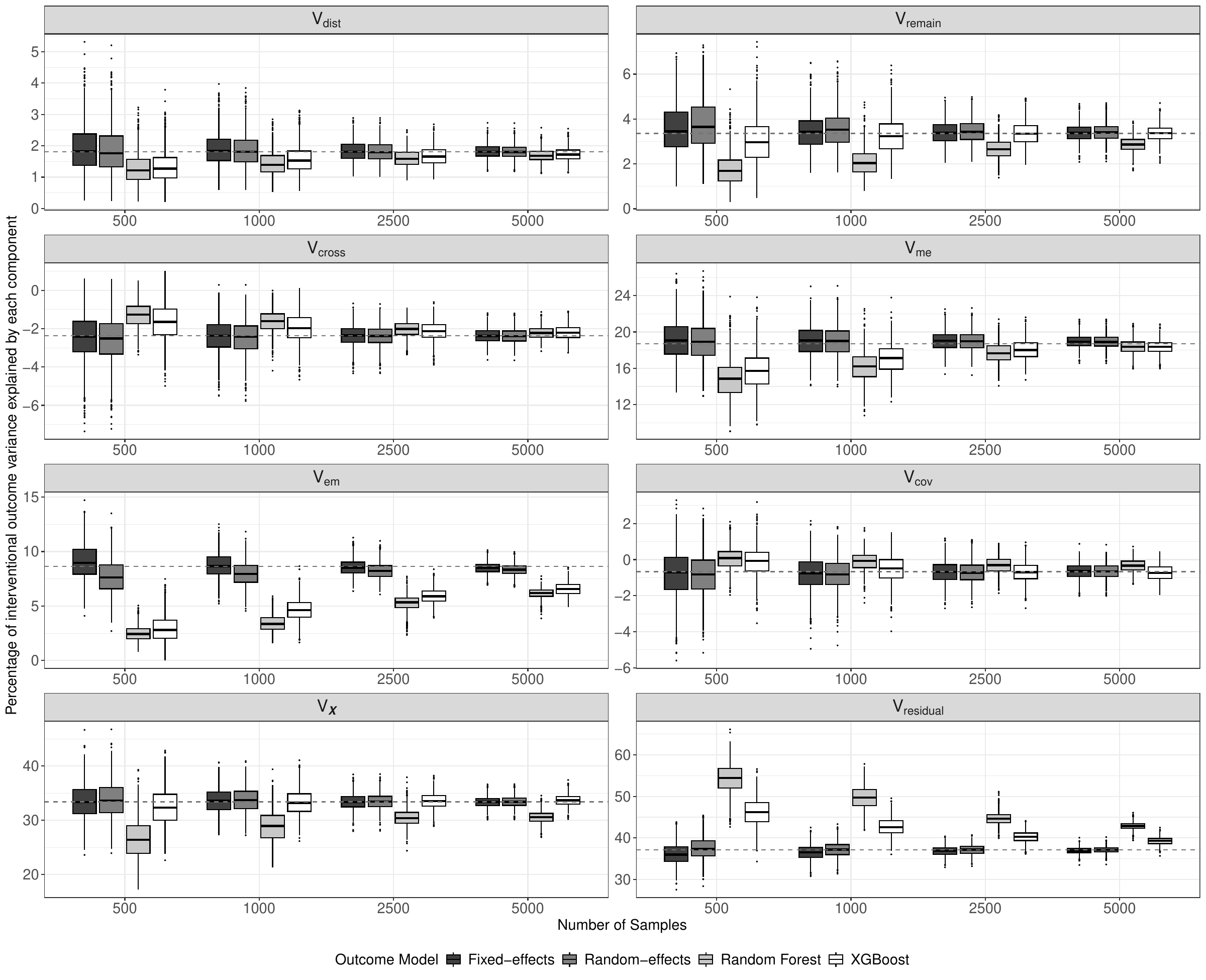}
    \caption{Sampling distributions of the estimators for a continuous outcome when $J = 5$.}
    \label{fig:hosp5_cont_res}
\end{figure}
\FloatBarrier

\begin{figure}[!htbp]
    \centering
    \includegraphics[width=1\linewidth]{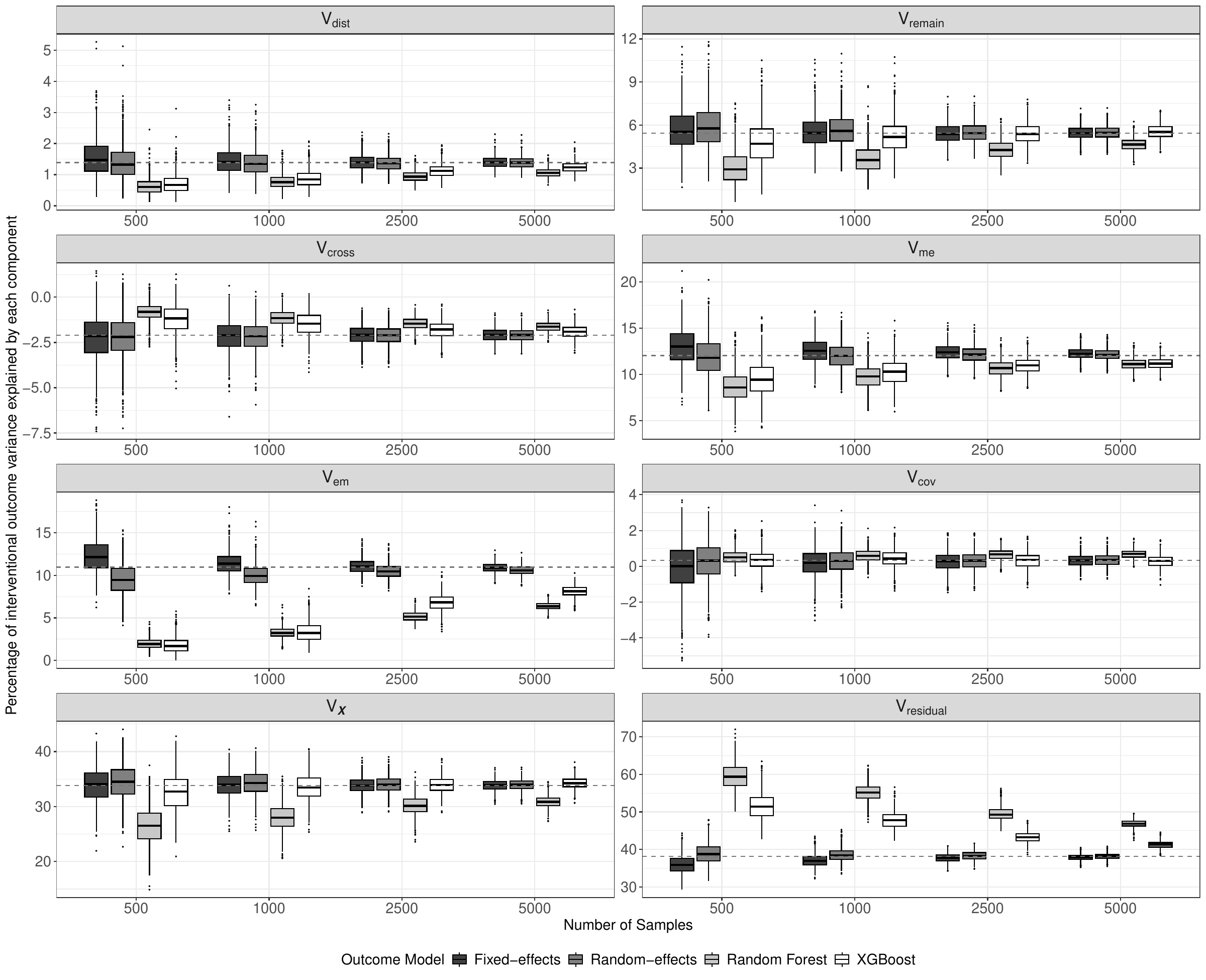}
    \caption{Sampling distributions of the estimators for a continuous outcome when $J = 10$.}
    \label{fig:hosp10_cont_res}
\end{figure}
\FloatBarrier

\section{Standard error assessment for fixed-effects estimators (\(J=5\))}
\label{sec:SE_MCSD_ratio_J5}
\begin{figure*}[!htbp]
\centering
\includegraphics[width=\textwidth]
{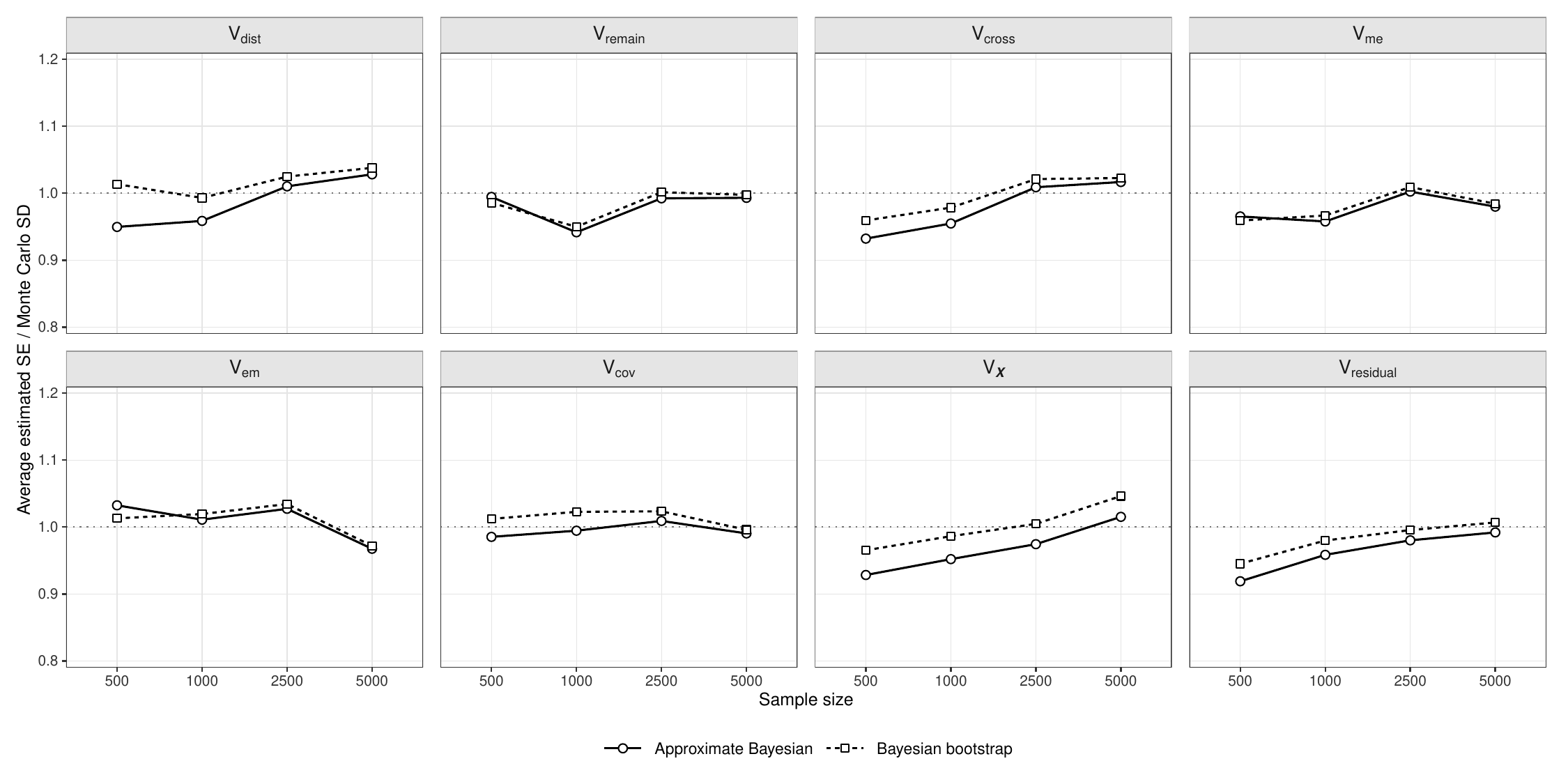}
\caption{Ratio of the average estimated standard error to the Monte
Carlo standard deviation for each variance component of a binary
outcome, using fixed-effects estimators with \(J=5\) care settings.
Each panel corresponds to one component, evaluated on its original
variance scale. Solid lines with open circles denote approximate
Bayesian parameter sampling, and dashed lines with open squares
denote the Bayesian bootstrap. The horizontal reference line at 1
indicates agreement. Both procedures were evaluated using the same
successful simulation replicates: 999, 1000, 1000, and 1000 at
\(n=500,1000,2500,\) and \(5000\), respectively.}
\label{fig:SE_MCSD_ratio_J5}
\end{figure*}
\FloatBarrier

\section{Descriptive statistics of the SEER Dataset}\label{sec:seer_eda}
Starting with the 8,500 patients in the cohort described by
Han et al.~\citep{han2024updated}, we sequentially excluded patients
with unknown metropolitan residence status (\(n=9\)), non-Hispanic
patients with unknown race (\(n=11\)), and patients with unknown
marital status (\(n=333\)). The exclusion of the Alaska Natives
registry yielded zero additional exclusions. We then excluded
75 non-Hispanic American Indian or Alaska Native patients because
this group was sparsely represented across registry areas. The
resulting analytic sample comprised 8,072 patients.
Table~\ref{tab:seer_cohort_characteristics} summarizes their demographic
and clinical characteristics by race and ethnicity.

\begin{table}[!htbp]
\caption{Demographic and clinical characteristics by race and ethnicity
in the analytic SEER cohort of 8,072 patients with cervical cancer.}
\label{tab:seer_cohort_characteristics}
\centering
\small
\setlength{\tabcolsep}{4pt}

\begin{tabular}{@{}lcccc@{}}
\toprule
\textbf{Characteristic}
& \textbf{\shortstack{Non-Hispanic\\White}}
& \textbf{\shortstack{Non-Hispanic\\Black}}
& \textbf{\shortstack{Hispanic\\(all races)}}
& \textbf{\shortstack{Non-Hispanic\\Asian or Pacific\\Islander}} \\
\midrule
Total, \(n\) & 3911 & 1259 & 2048 & 854 \\
\addlinespace
Registry, \(n\) (\%) & & & & \\
California            & 1114 (28.5) & 208 (16.5) & 1454 (71.0) & 555 (65.0) \\
Connecticut           & 155 (4.0)   & 44 (3.5)   & 49 (2.4)    & 6 (0.7) \\
Georgia               & 545 (13.9)  & 401 (31.9) & 79 (3.9)    & 31 (3.6) \\
Hawaii                & 15 (0.4)    & \textemdash & 10 (0.5)   & \textemdash \\
Iowa                  & 284 (7.3)   & 11 (0.9)   & 22 (1.1)    & 7 (0.8) \\
Kentucky              & 591 (15.1)  & 64 (5.1)   & 16 (0.8)    & 6 (0.7) \\
Louisiana             & 317 (8.1)   & 279 (22.2) & 20 (1.0)    & 10 (1.2) \\
New Jersey            & 499 (12.8)  & 230 (18.3) & 247 (12.1)  & 65 (7.6) \\
New Mexico            & 55 (1.4)    & \textemdash & 88 (4.3)   & \textemdash \\
Seattle (Puget Sound) & 221 (5.7)   & 12 (1.0)   & 35 (1.7)    & 60 (7.0) \\
Utah                  & 115 (2.9)   & 6 (0.5)    & 28 (1.4)    & 10 (1.2) \\
\addlinespace
Age at diagnosis, mean (SD)
& 52.05 (13.32) & 50.79 (14.15) & 49.75 (13.40) & 55.88 (13.56) \\
\addlinespace
Married, \(n\) (\%)
& 1672 (42.8) & 295 (23.4) & 757 (37.0) & 452 (52.9) \\
\addlinespace
Year of diagnosis, \(n\) (\%) & & & & \\
2000--2003 & 640 (16.4) & 239 (19.0) & 341 (16.7) & 121 (14.2) \\
2004--2006 & 547 (14.0) & 174 (13.8) & 281 (13.7) & 107 (12.5) \\
2007--2009 & 541 (13.8) & 188 (14.9) & 269 (13.1) & 109 (12.8) \\
2010--2012 & 543 (13.9) & 150 (11.9) & 277 (13.5) & 119 (13.9) \\
2013--2015 & 588 (15.0) & 176 (14.0) & 318 (15.5) & 145 (17.0) \\
2016--2018 & 696 (17.8) & 196 (15.6) & 336 (16.4) & 161 (18.9) \\
2019--2020 & 356 (9.1)  & 136 (10.8) & 226 (11.0) & 92 (10.8) \\
\addlinespace
Income, \(n\) (\%) & & & & \\
Below \$60,000
& 1336 (34.2) & 509 (40.4) & 329 (16.1) & 59 (6.9) \\
\$60,000--\$74,999
& 1334 (34.1) & 431 (34.2) & 1007 (49.2) & 297 (34.8) \\
\$75,000 or higher
& 1241 (31.7) & 319 (25.3) & 712 (34.8) & 498 (58.3) \\
\addlinespace
Cancer stage, \(n\) (\%) & & & & \\
IB2        & 483 (12.3)  & 145 (11.5) & 261 (12.7)  & 82 (9.6) \\
II         & 2038 (52.1) & 668 (53.1) & 1125 (54.9) & 462 (54.1) \\
III or IVA & 1390 (35.5) & 446 (35.4) & 662 (32.3)  & 310 (36.3) \\
\addlinespace
Metropolitan residence, \(n\) (\%)
& 3122 (79.8) & 1100 (87.4) & 1978 (96.6) & 825 (96.6) \\
\bottomrule
\end{tabular}

\par\smallskip
\begin{minipage}{\linewidth}
\footnotesize
\textit{Note.} Percentages are calculated within each racial and ethnic
group. Dashes indicate entries withheld to protect patient confidentiality
and prevent reconstruction of small counts.
\end{minipage}
\end{table}

\FloatBarrier

\section{Bayesian bootstrap sampling distribution of the estimators in the SEER study} \label{sec:AL_seer_sample_dist}
\begin{figure}[!htbp]
    \centering
    \includegraphics[width=1\linewidth]{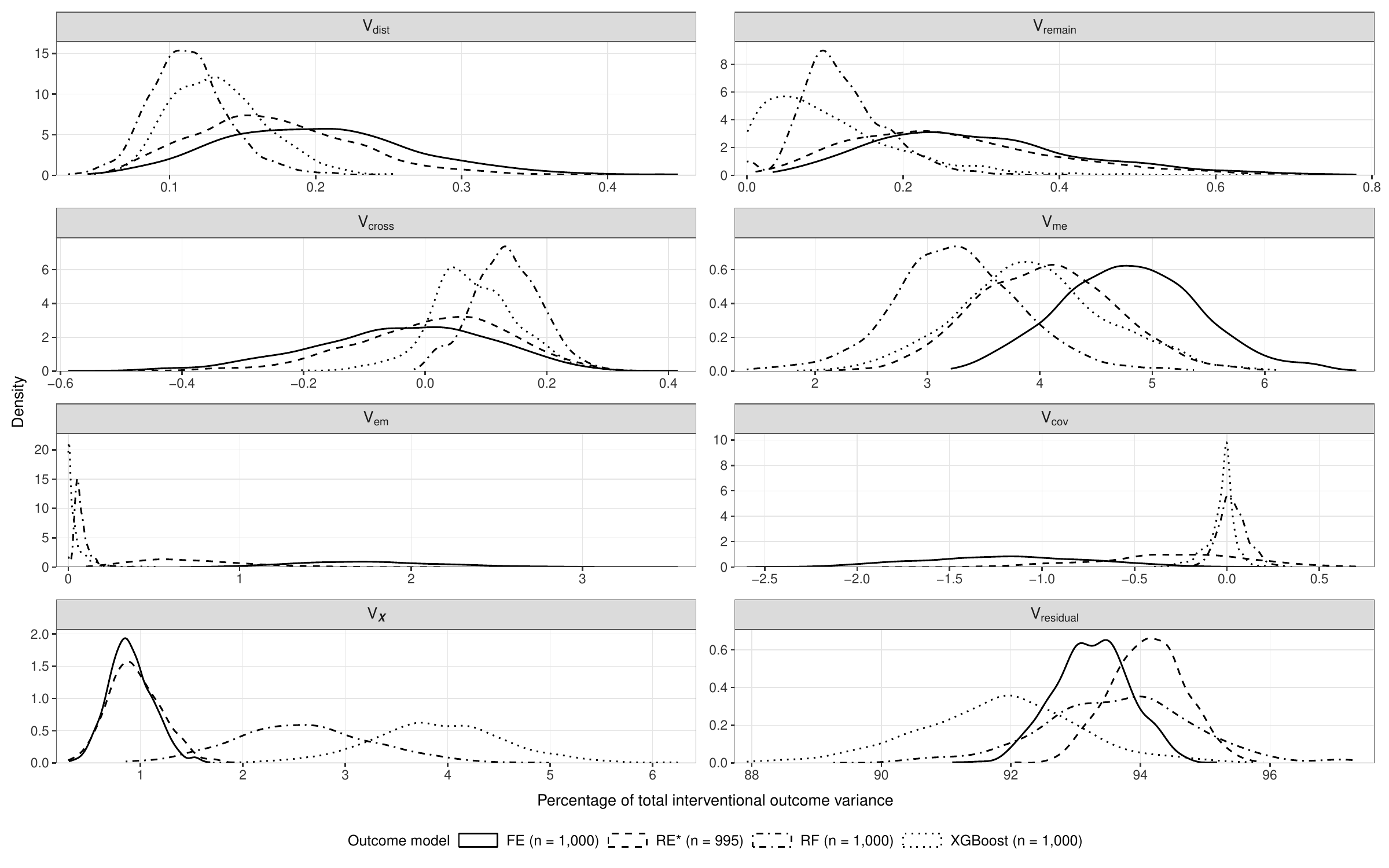}
    \caption{Bayesian bootstrap sampling distributions of the estimators.}
    \label{fig:seer_bootstrap_percent}
\end{figure}

%% file: bibliography.bib
@article{li2018balancing,
  author = {Li, Fan and Morgan, Kari Lock and Zaslavsky, Alan M.},
  title = {Balancing Covariates via Propensity Score Weighting},
  journal = {Journal of the American Statistical Association},
  year = {2018},
  volume = {113},
  number = {521},
  pages = {390--400},
  doi = {10.1080/01621459.2016.1260466}
}

@misc{seerRaceOriginRecode,
  author       = {{National Cancer Institute}},
  title        = {Race and {Hispanic} Ethnicity Changes},
  year         = {n.d.},
  howpublished = {Surveillance, Epidemiology, and End Results Program},
  url          = {https://seer.cancer.gov/seerstat/variables/seer/race_ethnicity/},
  note         = {Accessed September 24, 2026}
}

@article{hines2025variable,
  title={Variable importance measures for heterogeneous treatment effects},
  author={Hines, Oliver J and Diaz-Ordaz, Karla and Vansteelandt, Stijn},
  journal={Biometrics},
  volume={81},
  number={4},
  pages={ujaf140},
  year={2025},
  publisher={Oxford University Press}
}

@article{helpman2020disparities,
  title={Disparities in surgical management of endometrial cancers in a public healthcare system: a question of equity},
  author={Helpman, Limor and Pond, Gregory R and Elit, Lorraine and Anderson, Laura N and Seow, Hsien},
  journal={Gynecologic Oncology},
  volume={159},
  number={2},
  pages={387--393},
  year={2020},
  publisher={Elsevier}
}

@article{wolfstadt2019association,
  title={Association between socioeconomic deprivation and surgical complications in adults undergoing ankle fracture fixation: a population-based analysis},
  author={Wolfstadt, Jesse Isaac and Pincus, Daniel and Kreder, Hans J and Wasserstein, David},
  journal={Canadian Journal of Surgery},
  volume={62},
  number={5},
  pages={320},
  year={2019}
}

@article{shin2025treatment,
  title={Treatment effect heterogeneity and importance measures for multivariate continuous treatments},
  author={Shin, Heejun and Linero, Antonio and Audirac, Michelle and Irene, Kezia and Braun, Danielle and Antonelli, Joseph},
  journal={The Annals of Applied Statistics},
  volume={19},
  number={3},
  pages={1847--1867},
  year={2025},
  publisher={Institute of Mathematical Statistics}
}

@article{price2013racial,
  title={Racial/ethnic disparities in chronic diseases of youths and access to health care in the {United States}},
  author={Price, James H and Khubchandani, Jagdish and McKinney, Molly and Braun, Robert},
  journal={BioMed Research International},
  volume={2013},
  number={1},
  pages={787616},
  year={2013},
  publisher={Wiley Online Library}
}

@article{williams1995us,
  title={{US} socioeconomic and racial differences in health: patterns and explanations},
  author={Williams, David R and Collins, Chiquita},
  journal={Annual Review of Sociology},
  volume={21},
  number={1},
  pages={349--386},
  year={1995},
  publisher={Annual Reviews 4139 El Camino Way, PO Box 10139, Palo Alto, CA 94303-0139, USA}
}

@article{williams2010race,
  title={Race, socioeconomic status, and health: complexities, ongoing challenges, and research opportunities},
  author={Williams, David R and Mohammed, Selina A and Leavell, Jacinta and Collins, Chiquita},
  journal={Annals of the New York Academy of Sciences},
  volume={1186},
  number={1},
  pages={69--101},
  year={2010},
  publisher={Wiley Online Library}
}

@article{shen2025calibrated,
  title={A calibrated sensitivity analysis for weighted causal decompositions},
  author={Shen, Andy A and Visoki, Elina and Barzilay, Ran and Pimentel, Samuel D},
  journal={Statistics in Medicine},
  volume={44},
  number={5},
  pages={e70010},
  year={2025},
  publisher={Wiley Online Library}
}

@article{canedo2018racial,
  title={Racial/ethnic disparities in diabetes quality of care: the role of healthcare access and socioeconomic status},
  author={Canedo, Juan R and Miller, Stephania T and Schlundt, David and Fadden, Mary K and Sanderson, Maureen},
  journal={Journal of Racial and Ethnic Health Disparities},
  volume={5},
  number={1},
  pages={7--14},
  year={2018},
  publisher={Springer}
}

@article{george2017mortality,
  title={Mortality rate estimation and standardization for public reporting: {Medicare}’s {Hospital Compare}},
  author={George, Edward I and Ro{\v{c}}kov{\'a}, Veronika and Rosenbaum, Paul R and Satop{\"a}{\"a}, Ville A and Silber, Jeffrey H},
  journal={Journal of the American Statistical Association},
  volume={112},
  number={519},
  pages={933--947},
  year={2017},
  publisher={Taylor \& Francis}
}

@article{silber2016improving,
  title={Improving {Medicare}'s {Hospital Compare} mortality model},
  author={Silber, Jeffrey H and Satop{\"a}{\"a}, Ville A and Mukherjee, Nabanita and Ro{\v{c}}kov{\'a}, Veronika and Wang, Wei and Hill, Alexander S and Even-Shoshan, Orit and Rosenbaum, Paul R and George, Edward I},
  journal={Health Services Research},
  volume={51},
  pages={1229--1247},
  year={2016},
  publisher={Wiley Online Library}
}

@article{hartman2024evaluating,
  title={Evaluating medical providers in terms of patient health disparities: a statistical framework},
  author={Hartman, Nicholas and Dahlerus, Claudia},
  journal={Health Services and Outcomes Research Methodology},
  volume={24},
  number={4},
  pages={440--457},
  year={2024},
  publisher={Springer}
}

@article{chen2022causal,
  title={Causal mediation analysis decomposition of between-hospital variance},
  author={Chen, Bo and Lawson, Keith A and Finelli, Antonio and Saarela, Olli},
  journal={Health Services and Outcomes Research Methodology},
  pages={1--27},
  year={2022},
  publisher={Springer}
}

@article{van2024between,
  title={Between-hospital variation in indicators of quality of care: a systematic review},
  author={van der Linde, Margrietha and Salet, N{\`e}wel and van Leeuwen, Nikki and Lingsma, Hester F and Eijkenaar, Frank},
  journal={BMJ Quality \& Safety},
  volume={33},
  number={7},
  pages={443--455},
  year={2024},
  publisher={BMJ Publishing Group Ltd}
}

@article{varewyck2014shrinkage,
  title={On shrinkage and model extrapolation in the evaluation of clinical center performance},
  author={Varewyck, Machteld and Goetghebeur, Els and Eriksson, Marie and Vansteelandt, Stijn},
  journal={Biostatistics},
  volume={15},
  number={4},
  pages={651--664},
  year={2014},
  doi={10.1093/biostatistics/kxu019},
  publisher={Oxford University Press}
}

@misc{vanderlinde2026causal,
  author={van der Linde, M. and Yu, L. and Eijkenaar, F. and Bonsel, J. M.
          and Oude Groeniger, J. and van Steenbergen, L. N. and Beeres, D. T.
          and Eygendaal, D. and Oude Voshaar, A. H. M. and Lingsma, H. F.
          and Saarela, O.},
  title={Causal decomposition of health outcome disparities: Assessing the contribution of healthcare},
  year={2026},
  howpublished={SSRN preprint},
  doi={10.2139/ssrn.6073703},
  url={https://papers.ssrn.com/sol3/papers.cfm?abstract_id=6073703}
}

@article{tang2020constructing,
  author  = {Tang, Thai-Son and Austin, Peter C. and Lawson, Keith A.
             and Finelli, Antonio and Saarela, Olli},
  title   = {Constructing inverse probability weights for institutional
             comparisons in healthcare},
  journal = {Statistics in Medicine},
  year    = {2020},
  volume  = {39},
  number  = {23},
  pages   = {3156--3172},
  doi     = {10.1002/sim.8657}
}

@article{silber2010hospital,
  title={The hospital compare mortality model and the volume--outcome relationship},
  author={Silber, Jeffrey H and Rosenbaum, Paul R and Brachet, Tanguy J and Ross, Richard N and Bressler, Laura J and Even-Shoshan, Orit and Lorch, Scott A and Volpp, Kevin G},
  journal={Health Services Research},
  volume={45},
  number={5p1},
  pages={1148--1167},
  year={2010},
  publisher={Wiley Online Library}
}

@article{asch2021variation,
  title={Variation in {US} hospital mortality rates for patients admitted with {COVID-19} during the first 6 months of the pandemic},
  author={Asch, David A and Sheils, Natalie E and Islam, Md Nazmul and Chen, Yong and Werner, Rachel M and Buresh, John and Doshi, Jalpa A},
  journal={JAMA Internal Medicine},
  volume={181},
  number={4},
  pages={471--478},
  year={2021},
  publisher={American Medical Association}
}

@article{daignault2017doubly,
  title={Doubly robust estimator for indirectly standardized mortality ratios},
  author={Daignault, Katherine and Saarela, Olli},
  journal={Epidemiologic Methods},
  volume={6},
  number={1},
  pages={20160016},
  year={2017},
  publisher={De Gruyter}
}

@article{daignault2019causal,
  title={Causal mediation analysis for standardized mortality ratios},
  author={Daignault, Katherine and Lawson, Keith A and Finelli, Antonio and Saarela, Olli},
  journal={Epidemiology},
  volume={30},
  number={4},
  pages={532--540},
  year={2019},
  publisher={LWW}
}

@article{begg1984calculation,
  author  = {Begg, Colin B. and Gray, Robert},
  title   = {Calculation of polychotomous logistic regression parameters
             using individualized regressions},
  journal = {Biometrika},
  year    = {1984},
  volume  = {71},
  number  = {1},
  pages   = {11--18},
  doi     = {10.1093/biomet/71.1.11}
}

@article{jackson2021meaningful,
  title={Meaningful causal decompositions in health equity research: definition, identification, and estimation through a weighting framework},
  author={Jackson, John W},
  journal={Epidemiology},
  volume={32},
  number={2},
  pages={282--290},
  year={2021},
  publisher={LWW}
}

@article{jackson2026target,
  title={The target study: a conceptual model and framework for measuring disparity},
  author={Jackson, John W and Hsu, Yea-Jen and Greer, Raquel C and Boonyasai, Romsai T and Howe, Chanelle J},
  journal={Sociological Methods \& Research},
  volume={55},
  number={2},
  pages={405--458},
  year={2026},
  publisher={SAGE Publications Sage CA: Los Angeles, CA}
}

@article{chang2024importance,
  title={The importance of equity value judgments and estimator-estimand alignment in measuring disparity and identifying targets to reduce disparity},
  author={Chang, Ting-Hsuan and Nguyen, Trang Quynh and Jackson, John W},
  journal={American journal of epidemiology},
  volume={193},
  number={3},
  pages={536--547},
  year={2024},
  publisher={Oxford University Press}
}

@article{jackson2018decomposition,
  title={Decomposition analysis to identify intervention targets for reducing disparities},
  author={Jackson, John W and VanderWeele, Tyler J},
  journal={Epidemiology},
  volume={29},
  number={6},
  pages={825--835},
  year={2018},
  publisher={LWW}
}

@article{yu2025nonparametric,
  title={Nonparametric causal decomposition of group disparities},
  author={Yu, Ang and Elwert, Felix},
  journal={The Annals of Applied Statistics},
  volume={19},
  number={1},
  pages={821--845},
  year={2025},
  publisher={Institute of Mathematical Statistics}
}

@article{chen2020causal,
  title={Causal variance decompositions for institutional comparisons in healthcare},
  author={Chen, Bo and Lawson, Keith A and Finelli, Antonio and Saarela, Olli},
  journal={Statistical Methods in Medical Research},
  volume={29},
  number={7},
  pages={1972--1986},
  year={2020},
  publisher={SAGE Publications Sage UK: London, England}
}

@article{chen2023hierarchical,
  title={Hierarchical causal variance decomposition for institution and provider comparisons in healthcare},
  author={Chen, Bo and McAlpine, Kristen and Lawson, Keith A and Finelli, Antonio and Saarela, Olli},
  journal={Health Services and Outcomes Research Methodology},
  volume={23},
  number={4},
  pages={391--415},
  year={2023},
  publisher={Springer}
}

@book{pearl2009causality,
  title={Causality},
  author={Pearl, Judea},
  year={2009},
  publisher={Cambridge University Press}
}

@article{karvanen2024simulating,
  title={Simulating counterfactuals},
  author={Karvanen, Juha and Tikka, Santtu and Vihola, Matti},
  journal={Journal of Artificial Intelligence Research},
  volume={80},
  pages={835--857},
  year={2024}
}

@article{vanderweele2014effect,
  title={Effect decomposition in the presence of an exposure-induced mediator-outcome confounder},
  author={VanderWeele, Tyler J and Vansteelandt, Stijn and Robins, James M},
  journal={Epidemiology},
  volume={25},
  number={2},
  pages={300--306},
  year={2014},
  publisher={LWW}
}

@article{naimi2016mediation,
  title={Mediation analysis for health disparities research},
  author={Naimi, Ashley I and Schnitzer, Mireille E and Moodie, Erica EM and Bodnar, Lisa M},
  journal={American Journal of Epidemiology},
  volume={184},
  number={4},
  pages={315--324},
  year={2016},
  publisher={Oxford University Press}
}

@article{vanderweele2014causal,
  title={On the causal interpretation of race in regressions adjusting for confounding and mediating variables},
  author={VanderWeele, Tyler J and Robinson, Whitney R},
  journal={Epidemiology},
  volume={25},
  number={4},
  pages={473--484},
  year={2014},
  publisher={LWW}
}

@article{han2024updated,
  title     = {Updated trends in the utilization of brachytherapy in cervical cancer in the {United States}: a surveillance, epidemiology, and end-results study},
  author    = {Han, Kathy and Colson-Fearon, Darien and Liu, Zhihui Amy and Viswanathan, Akila N},
  journal   = {International Journal of Radiation Oncology*Biology*Physics},
  volume    = {119},
  number    = {1},
  pages     = {143--153},
  year      = {2024},
  doi       = {10.1016/j.ijrobp.2023.11.007},
  publisher = {Elsevier}
}

@article{probst2019hyperparameters,
  title={Hyperparameters and tuning strategies for random forest},
  author={Probst, Philipp and Wright, Marvin N and Boulesteix, Anne-Laure},
  journal={Wiley Interdisciplinary Reviews: Data Mining and Knowledge Discovery},
  volume={9},
  number={3},
  pages={e1301},
  year={2019},
  publisher={Wiley Online Library}
}

@misc{seerHome2025,
  author       = {{National Cancer Institute}},
  title        = {Surveillance, Epidemiology, and End Results ({SEER}) Program},
  year         = {2025},
  howpublished = {\url{https://seer.cancer.gov/}},
  note         = {Accessed: 2025-08-13},
}

@article{muller2005moderation,
  title={When moderation is mediated and mediation is moderated.},
  author={Muller, Dominique and Judd, Charles M and Yzerbyt, Vincent Y},
  journal={Journal of Personality and Social Psychology},
  volume={89},
  number={6},
  pages={852--863},
  year={2005},
  publisher={American Psychological Association}
}

@article{han2022restricted,
  title={Restricted mean survival time for survival analysis: a quick guide for clinical researchers},
  author={Han, Kyunghwa and Jung, Inkyung},
  journal={Korean Journal of Radiology},
  volume={23},
  number={5},
  pages={495--499},
  year={2022}
}

@article{valeri2023multistate,
  title={A multistate approach for the study of interventions on an intermediate time-to-event in health disparities research},
  author={Valeri, Linda and Proust-Lima, C{\'e}cile and Fan, Weijia and Chen, Jarvis T and Jacqmin-Gadda, H{\'e}l{\`e}ne},
  journal={Statistical Methods in Medical Research},
  volume={32},
  number={8},
  pages={1445--1460},
  year={2023},
  publisher={SAGE Publications Sage UK: London, England}
}

@article{khan2025marginal,
  title={Marginal and conditional importance measures from machine learning models and their relationship with conditional average treatment effect},
  author={Khan, Mohammad Kaviul Anam and Saarela, Olli and Kustra, Rafal},
  journal={arXiv preprint arXiv:2501.16988
        
        
        
        },
  year={2025}
}

@article{benmichael2025estimating,
  author  = {Ben-Michael, Eli and Feller, Avi and Kelz, Rachel and Keele, Luke},
  title   = {Estimating racial disparities in emergency general surgery},
  journal = {Journal of the Royal Statistical Society Series A: Statistics in Society},
  year    = {2025},
  volume  = {188},
  number  = {4},
  pages   = {1125--1148},
  doi     = {10.1093/jrsssa/qnae095},
  url     = {https://doi.org/10.1093/jrsssa/qnae095}
}

@misc{saarela2026graph,
  author={Saarela, Olli and Karvanen, Juha},
  title={Graph-based causal variance decompositions: When ``variance explained'' means causation},
  year={2026},
  howpublished={arXiv:2608.27140},
  doi={10.48550/arXiv.2608.27140},
  url={https://arxiv.org/abs/2608.27140}
}

@article{williamson2023general,
  title={A general framework for inference on algorithm-agnostic variable importance},
  author={Williamson, Brian D. and Gilbert, Peter B. and Simon, Noah R. and Carone, Marco},
  journal={Journal of the American Statistical Association},
  year={2023},
  volume={118},
  number={543},
  pages={1645--1658},
  doi={10.1080/01621459.2021.2003200}
}

@article{williamson2021nonparametric,
  title={Nonparametric variable importance assessment using machine learning techniques},
  author={Williamson, Brian D. and Gilbert, Peter B. and Carone, Marco and Simon, Noah},
  journal={Biometrics},
  year={2021},
  volume={77},
  number={1},
  pages={9--22},
  doi={10.1111/biom.13392}
}
